\documentclass[footinbib,twocolumn,showpacs,amsmath,amstex,amssymb,mathfonts,superscriptaddress,prx]{revtex4-2}
\usepackage{graphicx}
\usepackage{color}
\usepackage{bm}

\usepackage{graphicx}
\usepackage{color}
\usepackage{bm}
\usepackage{amsmath}
\usepackage{amssymb}
\usepackage{amsthm}
\usepackage{amsfonts}
\usepackage{multirow}
\usepackage{makecell}
\usepackage{float}
\usepackage{comment}
\usepackage{enumitem}
\usepackage{tipa}
\usepackage{mathtools}

\usepackage{bbm}

\begin{document}

\title{The Gaffnian and Haffnian: physical relevance of non-unitary CFT for incompressible fractional quantum Hall effect}
\author{Bo Yang} 
\affiliation{Division of Physics and Applied Physics, Nanyang Technological University, Singapore 637371.}
\affiliation{Institute of High Performance Computing, A*STAR, Singapore, 138632.}


\date{\today}
\begin{abstract}
We motivate a close look on the usefulness of the Gaffnian and Haffnian quasihole manifold (null spaces of the respective model Hamiltonians) for well-known gapped fractional quantum Hall (FQH) phases. The conformal invariance of these subspaces are derived explicitly from microscopic many-body states. The resultant CFT description leads to an intriguing emergent primary field with $h=2,c=0$, and we argue the quasihole manifolds are quantum mechanically well-defined and well-behaved. Focusing on the incompressible phases at $\nu=1/3$ and $\nu=2/5$, we show the low-lying excitations of the Laughlin phase are quantum fluids of Gaffnian and Haffnian quasiholes, and give a microscopic argument showing that the Haffnian model Hamiltonian is gapless against Laughlin quasielectrons. We discuss the thermal Hall conductance and shot noise measurements at $\nu=2/5$, and argue that the experimental observations can be understood from the dynamics within the Gaffnian quasihole manifold. A number of detailed predictions on these experimental measurements are proposed, and we discuss their relationships to the conventional CFT arguments and the composite fermion descriptions. 
\end{abstract}

\maketitle 

\section{Introduction}\label{intro}

The most important aspect of the fractional quantum Hall effect (FQHE) is the universal topological properties that manifest in experimental measurements\cite{prange}. Unlike symmetry protected topological phases, topological properties such as the Hall conductivity in FQHE are robust against any types of small perturbations. These properties arise from strong interaction between electrons confined not only in a two-dimensional manifold, but also in a single Landau level due to the strong magnetic field. The truncation of the Hilbert space (due to the smallness of the sample thickness and the magnetic length) at low temperature plays the crucial role here, leading to ground states with long range topological entanglement, fractionalisation of quasiparticle charges (i.e. anyons), nontrivial degeneracy of the quasihole manifold (i.e. non-Abelions), and chiral edge theories of at the boundary of the FQH fluids\cite{kitaev,wen,read,haldane,wen1}. 

Many exotic topological phases have been theoretically proposed, and their experimental realisations can potentially lead to robust storage and manipulations of quantum information\cite{nayak,simon}. The greatest challenge, however, is the existence of various different energy scales in realistic systems. Strictly speaking, in an effective description of a topological system all energy scales are set to either infinity (e.g. the incompressibility gap) or zero (e.g. the quasihole degeneracy), and we can denote the associated ``model Hamiltonian" as $\hat H_{\text{topo}}$. All the topological aspects are thus coming from a (possibly infinitely dimensional) sub-Hilbert space $\mathcal H_{\text{topo}}$ in which all states have \emph{zero energy}, while all states in the complementary sub-Hilbert space $\bar{\mathcal H}_{\text{topo}}$ have \emph{infinite energy}. For example in AKLT models with open boundary conditions, $\mathcal H_{\text{topo}}$ consists of the degenerate ground state manifold with different edge configurations\cite{aklt}. In the context of FQHE, $\mathcal H_{\text{topo}}$ consists of the ground state and all quasihole excitations that can be interpreted as edge excitations on a Hall manifold with a boundary. The existence of $\hat H_{\text{topo}}$ (not necessarily local) and a unique highest density state in $\mathcal H_{\text{topo}}$ (with electron density $\rho_{\text{max}}$) implies incompressibility for the FQH system, which is the necessary condition for the plateau of the Hall conductance in the presence of disorder. Experimentally, the incompressible state can be realised by local Hamiltonians adiabatically connected to $\hat H_{\text{topo}}$. 

When we smoothly go from $\hat H_{\text{topo}}$ to the realistic local Hamiltonian $\hat H_{\text{real}}$, we assume $\hat H_{\text{real}}$ is also incompressible at $\rho_{\text{max}}$, albeit with a finite gap. This implies a finite energy gap for all eigenstates with $\rho<\rho_{\text{max}}$, but there can be gapless excitations with $\rho>\rho_{\text{max}}$. In particular the state with $\rho_{\text{max}}$ does not necessarily have the lowest energy. For example in FQH systems, quasiholes can have either lower or higher energy depending on the disorder or the edge confinement potential. We can model the smooth deformation as follows:
\begin{eqnarray}\label{master}
\hat H=\left(1-\lambda\right)\hat H_{\text{topo}}+\lambda \hat H_{\text{real}}
\end{eqnarray}
A number of things can happen when $\lambda$ goes from $0$ to $1$, but in this work we will only focus on the following scenarios:
\begin{enumerate}[label={\bfseries P\arabic*}]
\item The subspaces $\mathcal H_{\text{topo}}\left(\lambda\right)$ and $\bar{\mathcal H}_{\text{topo}}\left(\lambda\right)$ evolve adiabatically and are completely gapped for the entire range of $\lambda$; there is no level crossing between any two states from the different subspaces.
\begin{enumerate}
\item All states in $\mathcal H_{\text{topo}}$ remain degenerate in the thermodynamic limit at $\lambda=1$.
\item Degeneracy of states in $\mathcal H_{\text{topo}}$ gets lifted, developing a finite bandwidth $\Delta_{\text{topo}}$ at $\lambda=1$.
\end{enumerate}
\item Level crossing occurs between some states from the two subspaces, but the quantum sector of the highest density state $|\psi_{\rho_{\text{max}}}\rangle$ is adiabatically connected for $\lambda$ between $0$ and $1$, which remains incompressibile.
\item  A finite energy gap opens up within $\mathcal H_{\text{topo}}$ at $\lambda=1$ in the thermodynamic limit, with a new gapped ground state $|\psi_{\rho_0}\rangle$ at $\rho_0<\rho_{\text{max}}$. This implies no level crossing between $\mathcal H_{\text{topo}}$ and $\bar{\mathcal H}_{\text{topo}}$ in the quantum sector of $|\psi_{\rho_0}\rangle$ for $\lambda$ between $0$ and $1$.
\end{enumerate}

For simplicity we only consider a single species of spinless fermions here, and the quantum sectors are labeled by the total number of electrons $N_e$, and the total angular momentum $M$ on the disk geometry. The topological properties of $\hat H_{\text{real}}$ are thus completely determined by the behaviours listed above. If level crossing occurs in the quantum sector of the highest density state, then $\hat H_{\text{topo}}$ and $\hat H_{\text{real}}$ are not topologically related in every sense. Statement $1$ is definitely possible if $\hat H_{\text{real}}$ is a small perturbation from $\hat H_{\text{topo}}$. If Statement $1(a)$ is the case, then $\hat H_{\text{topo}}$ and $\hat H_{\text{real}}$ are completely equivalent topologically. However, we should not take this for granted, as it is not fundamentally forbidden for Statement $1(b)$ to be true. If that is the case, then one would expect certain physical properties considered universal at $\hat H_{\text{topo}}$ will be lost with the realistic Hamiltonian. These could include universal edge behaviours as predicted by the chiral Luttinger liquid theory\cite{chang,wen2,read2}, as well as the non-Abelian braiding of quasiholes\cite{read5} which fundamentally depends on the degeneracy of the quasihole manifold.

In numerical calculations, the most common behaviour observed is actually Statement $2$, especially for FQH in higher LLs and for the non-Abelian FQH states. While this could well be the finite size effect when realistic Hamiltonians are used, we should also take the possibility seriously that some level crossing could occur in the thermodynamic limit, while the ground state remains incompressible. The robustness of the Hall conductivity plateau only requires incompressibility and thus the adiabatic continuity of $|\psi_{\rho_{\text{max}}}\rangle$, and this does not automatically imply the robustness of any or all of the other topological properties of the FQH state. If level crossing occurs despite the highest density state being adiabatically connected, one could argue that $\hat H_{\text{topo}}$ and $\hat H_{\text{real}}$ no longer belong to the same universality class. Indeed, we would expect gapless boundary states to develop at the interface of the two Hamiltonian (or at certain values of $\lambda$ when level crossing occurs). This could the be underlying difference between some of the known distinct FQH phases\cite{yang1}, though in this case all topological properties of the highest density state (i.e. the ground state) are equivalent for the two phases.

Statement 3 and Statement 2 are not mutually exclusive, though Statement 3 implies a topological phase transition even for the ground state. Let $|\psi^{\alpha_1}_{\rho_{\text{max}}}\rangle$ and $|\psi^{\alpha_2}_{\rho_0}\rangle$ be the two global ground states at $\lambda=0$ and $\lambda=1$ respectively with $\rho_0<\rho_{\text{max}}$, and $\alpha_1,\alpha_2$ are indices of quantum numbers from symmetries common to both $\hat H_{\text{topo}}$ and $\hat H_{\text{real}}$ (e.g. angular momentum $M$). If $|\rho_0-\rho_{\text{max}}|$ is sub-extensive (with respect to $N_e$), we still expect the two topological phases to occur at the same filling factor or the Hall plateau in the thermodynamic limit, and this could be the case for FQH phases that differ by the topological shift (related to Hall viscosity\cite{zee,rezayi}). It is also interesting to consider the case that $|\alpha_1-\alpha_2|$ is extensive, which implies both states can be (meta)stable for $0<\lambda<1$, since disorder will not be able to mix the two states in the thermodynamic limit. In analogy to the case of spontaneous symmetry breaking, we will have a spontaneous ``topology" breaking, if the energies of $|\psi^{\alpha_1}_{\rho_{\text{max}}}\rangle$ and $|\psi^{\alpha_2}_{\rho_0}\rangle$ are close as compared to disorder or temperature. A familiar example is the competition between the Pfaffnian and anti-Pfaffnian phase at half-filling, when the two-body interaction (which is particle-hole symmetric) dominates\cite{yang2,rosenow,fisher,rezayi2}. On the disk, the ground states of the two phases live in two angular momentum sectors that are infinitely apart in the thermodynamic limit, because the two phases have different topological shifts ($s=-2$ for Pfaffian and $s=+2$ for anti-Pfaffnian). Thus local disorders alone will not be able to lift the degeneracy of the two states.

In this paper, we will discuss the physical properties of two rather special FQH candidates, the Gaffnian state and the Haffnian state, in the context of the three Statements above. These two states are special, because both from the effective theory description (e.g. the conformal field theory, or CFT) and the microscopic model perspective (e.g. three-body local model Hamiltonians), the Hilbert spaces ${\mathcal H}_{\text{topo}}$ and $\bar{\mathcal H}_{\text{topo}}$ can be unambiguously defined. However, the corresponding CFT models are non-unitary and/or irrational\cite{read2,hermanns,simon3}. It is thus conjectured that there can be no \emph{local} Hamiltonians to give $\bar{\mathcal H}_{\text{topo}}$ a finite energy gap, while keeping all states in ${\mathcal H}_{\text{topo}}$ strictly degenerate (at zero energy). Thus the Gaffnian and Haffnian states are considered to be related to some critical gapless phases in FQH systems\cite{read6}, and it is generally dismissed as physically irrelevant to gapped FQH phases. We would like to examine these notions in more details in this work.

The organisation of the paper will be as follows: In Sec.~\ref{cft}, we give an overview of the well-known effective description of conformal invariance of the FQH systems, and give a rather different derivation of conformal invariance for the FQH quasihole subspace from the microscopic many-body states; in Sec.~\ref{lambda0} we apply the microscopic derivation of conformal invariance to the Laughlin state and the model states from three-body pseudopotential interactions (including the Moore-Read, Gaffnian and Haffnian states), leading to a number of interesting observations about the nature of their quasihole manifold; in Sec.~\ref{excitations} we motivate the physical relevance of the Gaffnian and Haffnian states to gapped FQH phases, showing that they are model states for elementary excitations in the Laughlin phases. Interestingly, this also leads to a semi-rigorous microscopic argument on why the Haffnian model Hamiltonian is gapless in the thermodynamic limit, though the same arguments do not seem to apply to the Gaffnian model Hamiltonian; in Sec.~\ref{realistic} we move onto more realistic interactions, and argue that using the Gaffnian and Haffnian quasihole subspaces, a number of experimental results can be explained, and some more detailed predictions can be made. These include the physics related to the thermal Hall effect and the shot noise/quasihole tunnelling experiments; in Sec.~\ref{summary}, we give a summary with further discussions. In particular we will summarise a number of detailed predictions from the analysis in this work, that can be directly related to experiments.

\begin{table}[h!]
\centering
\begin{tabular}{ |c|c|} 
 \hline
 Topological Hilbert space & $\mathcal H_{\text{topo}}^I$ \\ 
 \hline
  Full Hilbert space&$\mathcal H_{\text{topo}}^I\otimes\bar{\mathcal H}_{\text{topo}}^I$\\ 
 \hline
 Bandwidth of $\mathcal H_t^I$ & $\Delta_{\text{topo}}^I$\\
 \hline
 Highest density state in $\mathcal H_t^I$ & $|\psi^I_{\rho_{\text{max}}}\rangle$\\
 \hline
 Electron number & $N_e$\\
 \hline
 Total angular momentum & $M$\\
 \hline
  A many-body quantum state in $\mathcal H_t^I$ &$\psi_{k,I}$\\
\hline
 A many-body primary state in $\mathcal H_t^I$ & $\psi_{h_\alpha,I}$\\
 \hline
 The holomorphic part of $\psi_{k,I}$&$\phi_{k,I}$ or $|\phi_{k,I}\rangle$\\
 \hline
 A descendant state at level $N$ in $\mathcal H_t^I$&$|\phi_{h_\alpha,I}^{\left(N\right)}\rangle, |\phi_{h_I}^{\left(N\right)}\rangle$\\
 \hline
 A second primary state with $h=2,c=0$&$|\tau_I\rangle$\\
 \hline
\end{tabular}
\caption{Various notations used in this paper.}
\label{t0}
\end{table}

\section{CFT description of FQHE}\label{cft}

We will now specialise to the FQH systems, so that $\mathcal H_{\text{topo}}$ is the Hilbert space of the FQH ground state and its quasihole excitations (degenerate states that are less dense than the ground state), while $\bar{\mathcal H}_{\text{topo}}$ is the Hilbert space of all gapped excitations, including the neutral and quasielectron excitations. On the disk or cylinder geometry where edges are present, each quasihole state can be reinterpreted as an edge excitations. This is because the insertion of magnetic fluxes (to create quasiholes) pushes electrons to the boundary, even if the insertion is deep in the bulk. For quasiholes created deep in the bulk, however, they correspond to edge excitations with very large momenta. Thus those states are not necessarily important for the low energy, long wavelength limit of the edge theory.

Using the disk geometry, an important way of analysing many FQH phases is to use the conformal field theory (CFT), which is particularly useful for two-dimensional critical systems with conformal symmetry\cite{cft,textbook}. The FQH systems are insulators that are gapped in the bulk. However for a quantum Hall droplet with a boundary (which is also a more realistic scenario in the experiments), it is indeed a gapless system due to the gapless edge excitations. For all energy scales smaller than the bulk gap, we can thus treat the edge dynamics as a one-dimensional chiral system\cite{read2}. One can show that this one-dimensional system is conformally invariant in the $1+1$ space-time, if it is \emph{maximally chiral}: all excitations travel at a common velocity $v$. Formally, for any local operator $\hat O$ at the edge, we need to have:
\begin{eqnarray}\label{ov}
\hat O\left(x,t\right)=\hat O\left(x-vt\right)
\end{eqnarray}
where $x$ is the periodic spatial coordinate at the edge, and $t$ is time. The CFT description is thus an effective theory for the Hilbert space of $\mathcal H_{\text{topo}}$ only. The states in $\mathcal H_{\text{topo}}$ are only degenerate in the limit of $v\rightarrow 0$. For any non-zero value of $v$, however, $\mathcal H_{\text{topo}}$ satisfies conformal symmetry with the assumption of Eq.(\ref{ov}).

Thus if we project into the Hilbert space of $\mathcal H_{\text{topo}}$, assuming the linear spectrum and Eq.(\ref{ov}) can be realised with certain physical Hamiltonian, then all edge excitations can be described by an effective model satisfying conformal symmetry. Such models can also be analysed and understood via the elegant machinery of CFT. There is thus a natural bulk-edge correspondence, because each edge mode can also be understood as a quasihole excitation (i.e. conformally mapped to the spherical geometry as a bulk excitation in $\mathcal H_{\text{topo}}$). It has been discovered for some FQH phases, that the microscopic models wavefunctions on the disk geometry for all states in $\mathcal H_{\text{topo}}$ are give by the correlators or conformal blocks of certain CFT models\cite{read3,read,read5}. Such models thus have to encode some information about the edge dynamics of the same FQH phases\cite{haldane2,anushya}.

We now work under the assumption that every state in $\mathcal H_{\text{topo}}$ can be written as the correlator of a specific CFT model $\mathcal M_{\mathcal H_{\text{topo}}}$. This can be verified microscopically for a number of FQH models. It is thus pertinent to ask about the relationship between $\mathcal H_{\text{topo}}$ and the Hilbert space of $\mathcal M_{\mathcal H_{\text{topo}}}$. The CFT correlators that map to every microscopic state in $\mathcal H_{\text{topo}}$ only consists of primary fields in $\mathcal M_{\mathcal H_{\text{topo}}}$. Each quasihole corresponds to one primary field in the correlator with coordinates $\eta_i=x_i+iy_i$, where the subscript is the quasihole index, and $x,y$ are the real space coordinates of the two-dimensional disk. The degeneracy of the multi-quasihole states after fixing their locations, which is important to the non-Abelian properties of the FQH phase, is also determined by the distinct correlators from the fusion rules of those primary fields. In fact for all known FQH phases with exact model Hamiltonians (so that the ground state and quasiholes are well-defined zero energy states), there is a one-to-one mapping of every state in $\mathcal H_{\text{topo}}$ to every possible correlators involving only the primary fields. These states are wavefunctions of the locations of the electrons and the quasiholes. The number of primary fields in the correlator corresponds to the number of electrons and quasiholes in the many-body wavefunction. Thus from this perspective, the Hilbert space of $\mathcal M_{\mathcal H_{\text{topo}}}$ is much larger than $\mathcal H_{\text{topo}}$, since no descendant fields are involved in the construction of states in $\mathcal H_{\text{topo}}$. Note that this perspective is the CFT description of the FQH bulk properties, which is also valid on geometries with no boundaries, such as sphere or torus. For some hierarchical states with no known model Hamiltonians (but may have approximate ones\cite{sreejith1}), there are attempts to write their many-body wavefunctions as CFT correlators involving descendant fields\cite{hansson1,kunyang}. Here we restrict ourselves to only ones with exact model Hamiltonians.

If we focus on the edge excitations as a dynamical system with linear dispersion (with the same Hilbert space $\mathcal H_{\text{topo}}$), then the CFT model $\mathcal M_{\mathcal H_{\text{topo}}}$ is an effective theory so far with no rigorous microscopic derivation. The coordinates of the effective theory are no longer real space coordinates, but holomorphic and anti-holomorphic coordinates $z=x+ivt, \bar z=x-ivt$. Here $x$ is the periodic coordinate along the edge of the disk, and $t$ is the time, which we can take along the radial direction. The Hilbert space in $\mathcal M_{\mathcal H_{\text{topo}}}$ is thus generated by conformal generators $\hat{\mathcal L}_n$ satisfying the Virasoro algebra:
\begin{eqnarray}\label{virasoro}
 [\hat {\mathcal L}_n,\hat {\mathcal L}_m]=\left(n-m\right)\hat {\mathcal L}_{m+n}+\frac{c}{12}n\left(n^2-1\right)\delta_{m+n,0}
 \end{eqnarray}
 where $c$ is the central charge of the CFT model. The natural Hamiltonian is proportional to $\hat{\mathcal L}_0$, which is the dilation operator and thus the translation along the time (i.e. radial) direction. We thus expect all states in $\mathcal H_{\text{topo}}$ to be mapped to the primary and descendant fields in $\mathcal M_{\mathcal H_{\text{topo}}}$, which are eigenstates of $\hat{\mathcal L}_0$ in the CFT description. 
 
\begin{figure}
\includegraphics[width=\linewidth]{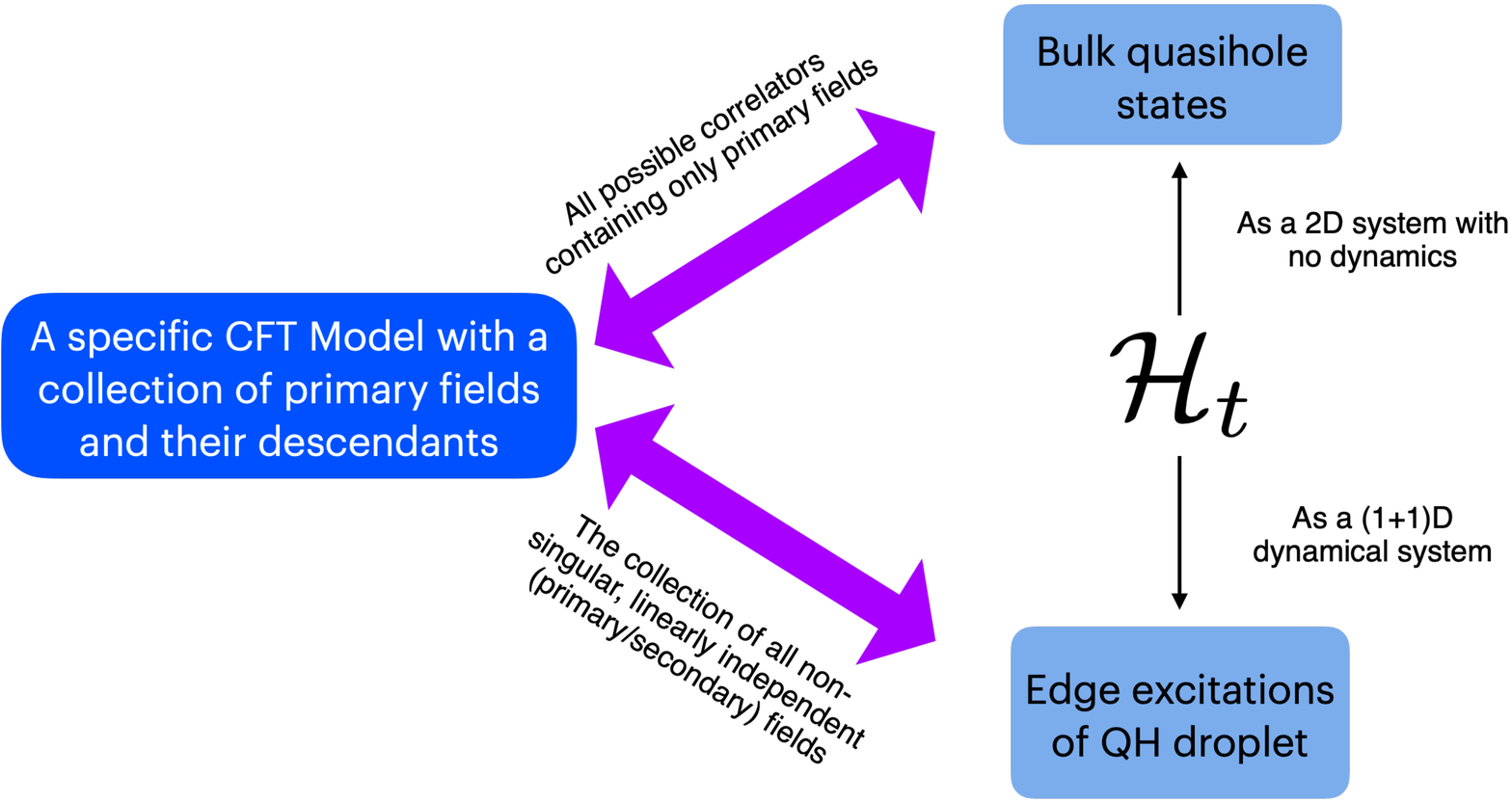}
\caption{Relationship between the effective CFT models and the microscopic many-body states, from the bulk and edge perspectives.}
\label{fig1}
\end{figure}

Thus while it is microscopically equivalent for $\mathcal H_{\text{topo}}$ to be treated as bulk quasihole excitations, or as edge excitations at the boundary of the quantum Hall droplet, the corresponding CFT descriptions are rather distinctive (see Fig.(\ref{fig1})). As quasiholes, $\mathcal H_{\text{topo}}$ is mapped to primary field correlators in the two-dimensional manifold (no dynamical information); as edge excitations, $\mathcal H_{\text{topo}}$ is mapped to all primary fields and descendant fields in one dimension, with a well-defined ``conformal Hamiltonian". This bulk-edge correspondence, or the equivalence of the two descriptions, should in general not be true for any arbitrarily defined subspace. It is fundamentally due to the intrinsic topological or algebraic structures of $\mathcal H_{\text{topo}}$. The one-to-one mapping of the states in $\mathcal H_{\text{topo}}$ and the primary/descendant fields of the edge CFT can also be quite non-trivial, which is a particularly important issue when $\mathcal M_{\mathcal H_{\text{topo}}}$  is nonunitary or irrational.

\subsection{The nonunitary and irrational CFT models}  

The CFT description and the bulk-edge correspondence seem to work quite well, if the CFT model $\mathcal M_{\mathcal H_{\text{topo}}}$ is rational and unitary. In these cases, $\mathcal M_{\mathcal H_{\text{topo}}}$ only contains a finite number of primary fields, and the norm of all primary and descendant fields in the model, as defined in CFT, are non-negative. The description becomes subtle when $\mathcal M_{\mathcal H_{\text{topo}}}$ is non-unitary and or irrational, which is relevant to the Gaffnian state\cite{simon2} (non-unitary) and the Haffnian state\cite{ardonne,read6} (non-unitary and irrational) we will focus on for the main part of this work.

It is generally argued that the nonunitary/irrational CFT models cannot be physical models in describing the dynamics of the conformally invariant one-dimensional edge systems\cite{read5,simon3}. This is because nonunitary CFT models contain fields with negative norm and thus diverging correlation functions at the edge. Irrational CFT models imply an infinite number of primary fields which seems unnatural. On the torus geometry, this implies the ground state degeneracy (the number of highest density states in $\mathcal H_{\text{topo}}$) is infinity in the thermodynamic limit, making it unlikely to describe a gapped bulk phase. The proposed resolution is that $\mathcal H_{\text{topo}}$ alone cannot describe the physical edge dynamics in these cases. The low lying gapless excitations have to include states from $\bar{\mathcal H}_{\text{topo}}$, thus the bulk gap has to close in the thermodynamic limit, so that the mapping to non-unitary/irrational CFT models will no longer hold.

To get a fuller picture of the scenario, we first note that $\mathcal H_{\text{topo}}$ is well-defined with nonunitary/irrational CFT model on the disk geometry or local Hamiltonians, since the latter is gapless only in the thermodynamic limit, which is an asymptotic behaviour. The one-to-one mapping of the null space to the primary field correlator, as well as to all of the primary/descendant fields, can be established. There is thus no ambiguity in defining $\mathcal H_{\text{topo}}$ as the null space of a particular $\hat H_{\text{topo}}$ with an infinite gap to $\bar{\mathcal H}_{\text{topo}}$, as long as we do not require $\hat H_{\text{topo}}$ to be local. Thus the arguments about the inability to define $\mathcal H_{\text{topo}}$ as the gapped null space can only be applied to local Hamiltonians, which we will focus in the next section. On the other hand, it is not clear if the mapping between quantum states in $\mathcal H_{\text{topo}}$ and the primary/descendant fields in $\mathcal M_{\mathcal H_{\text{topo}}}$ goes beyond state counting. In particular, all states in $\mathcal H_{\text{topo}}$ have a well-defined, positive definite quantum mechanical norm, even when their counterparts can have negative norms defined in CFT. The quantum mechanical quasihole correlation functions in $\mathcal H_{\text{topo}}$ also decay as a function of the distance between them, while in the CFT description the correlation diverges due to the negative conformal dimensions\cite{bernevig1,bernevig2}.

This apparent inconsistency implies it is important to understand what physical aspects of $\mathcal H_{\text{topo}}$ can be captured by the corresponding CFT model, in addition to the state counting and the linear spectrum. The linear spectrum from maximal chirality is also not intrinsic to $\mathcal H_{\text{topo}}$, but rather from a putative effective Hamiltonian on states in the restricted Hilbert space (the null space), that are expected to be only physically relevant in the long wavelength limit of the edge system from the confining potential. Given that the only constraint or assumption here is the conformal invariance of $\mathcal H_{\text{topo}}$, the resolution could ultimately be about the derivation of effective CFT theory from the microscopic details of the Hilbert space. 

\subsection{Microscopic derivation of conformal invariance of $\mathcal H_{\text{topo}}$}

We will show a rather crude attempt here that nevertheless leads to a number of interesting results relevant to the main focus of this work, but leave a detailed discussion for future works. To start from the familiar grounds, let us focus on the lowest Landau level (LLL) first, though the main results derived here applies to any Landau level as they should be. Every state in $\mathcal H_{\text{topo}}$ is a many-body wavefunction with holomorphic variables $z_i=x_i+iy$, where the subscript $i$ is the electron index. We thus denote $\mathcal H_{\text{topo}}=\{\psi_k\left(z_1,z_2,z_3,\cdots\right)\}$, and the states can have any number of electrons. Without loss of generality, we can fix the total number of electrons to be $N_e$, assuming the electron number as a good quantum number. In the LLL, the many-body wavefunctions are given by:
\begin{eqnarray}\label{states}
\psi_k\left(z_1,z_2\cdots z_{N_e}\right)=\phi_k\left(z_1,z_2\cdots z_{N_e}\right)e^{-\frac{1}{4}\sum_iz_iz_i^*}
\end{eqnarray}
where $\phi_k\left(z_1,z_2\cdots z_{N_e}\right)$ is holomorphic in its variables, as a linear combination of the monomial basis. We can thus write $\phi_k=\sum_\lambda c_{k\lambda}m_\lambda$, where $m_\lambda=\text{Asy}\left(z_1^{n_{1\lambda}}z_2^{n_{2\lambda}}\cdots z_{N_e}^{n_{N_e\lambda}}\right)$. The antisymmetrisation $\text{Asy}$ is over the electron indices. In addition, we assume rotational invariance on the disk geometry, thus the total angular momentum $M_\lambda$ is also a good quantum number, with $M_\lambda=\sum_in_{i\lambda}$. Each state physically represents a quantum Hall droplet, the size of which is given by the highest power of $z_i$ in the monomial basis of  $\phi_k\left(z_1,z_2\cdots z_{N_e}\right)$. The Gaussian factor in Eq.(\ref{states}) is not important, so we will denote states in $\mathcal H_{\text{topo}}$ with $\phi_k$.

It is also useful to have the second quantised representation of $\phi_k$, so we can denote each monomial as follows:
\begin{eqnarray}
 m_\lambda=\text{Asy}\left(z_1^{n_{1\lambda}}z_2^{n_{2\lambda}}\cdots z_{N_e}^{n_{N_e\lambda}}\right)\sim c_{n_{1\lambda}}^\dagger c_{n_{2\lambda}}^\dagger\cdots c_{n_{N_e\lambda}}^\dagger|\text{vac}\rangle\qquad
 \end{eqnarray}
 where $c_i^\dagger$ is the electron creation operator in the single particle orbital indexed by $i$(i.e. $z^i$), satisfying the anticommutation relations $\{c_i,c_j^\dagger\}=\delta_{ij},\{c_i,c_j\}=\{c_i^\dagger,c_j^\dagger\}=0$. We can thus represent $m_\lambda$ with a binary string. Each digit from left to right corresponds to a single particle orbital from the center to the edge of the disk (i.e. indexed by the power of $z$). As an example with two electrons, we have $\left(z_1-z_2\right)^3=\left(z_1^3-z_2^3\right)-3\left(z_1^2z_2-z_1z_2^2\right)=|1001000\cdots\rangle - 3|0110000\cdots\rangle$. One should note the second quantised representation is applicable to FQH in any LLs. All the results in this section can be derived from the second quantised representation, and they do not require the holomorphic wavefunctions that are specific to the LLL.

Given the conformal invariance of $\mathcal H_{\text{topo}}$, we should be able to identify one or more states $\phi_{h_\alpha}\in\mathcal H_{\text{topo}}$ as the primary states, indexed by $\alpha$  (as we will properly define later). These are states that are analogous to the ``primary fields" in CFT. All other states in $\mathcal H_{\text{topo}}$ can be generated from the primary states by operators satisfying the familiar Virasoro algebra. To that end, we need to properly define the Virasoro generators acting on $\phi_k$. Given that the classical version $\hat l_{-n}=\sum_i z_i^{n+1}\partial_{z_i}$ satisfies Eq.(\ref{virasoro}) with $c=0$, we can  define the following second quantised operators analog with $n\ge 0$:
\begin{eqnarray}
&&\hat L_{-n}=\sum_{k=0}^\infty f_{k+n,k}\cdot k\hat c^\dagger_{k+n}\hat c_k\label{vv1}\\
&&\hat L_{n}=\sum_{k=0}^\infty f_{k,k+n}\cdot \left(k+n\right)\hat c^\dagger_{k}\hat c_{k+n}\label{vv2}
\end{eqnarray}
The function $f_{k_1,k_2}$ comes from the single particle state normalisation that depends on the geometry. For example on the disk geometry, $f_{k_1,k_2}=\sqrt{k_1!/k_2!}$. It is easy to check that the Virasoro algebra is satisfied between $\hat L_m,\hat L_n$ if $n\cdot m\ge 0$. The commutation relation between the positive and negative modes is a bit more subtle:
\begin{eqnarray}
&&[\hat L_{m},\hat L_{-n}]=\left(n-m\right)\hat L_{m+n}+\hat C_{m,n}\label{vv3}\\
&&\hat C_{m,n}=\begin{cases}
\sum\limits_{k=0}^{m-1}f_{k,k+\Delta}\cdot\left(k-m\right)\left(k+\Delta\right)\hat c_k^\dagger\hat c_{k+\Delta}&n\ge m\qquad\\
\sum\limits_{k=0}^{n-1}f_{k+\Delta,k}\cdot k\left(k-n\right)c^\dagger_{k+\Delta}\hat c_k&n\le m\qquad
\end{cases}
\end{eqnarray}
with $m,n\ge 0, \Delta=|m-n|$. Thus the Virasoro algebra is not explicitly obeyed by the additional term $\hat C_{m,n}$. 

On the other hand, we expect the conformal symmetry to be satisfied only in the thermodynamic limit, and the Virasoro algebra to be obeyed only within $\mathcal H_{\text{topo}}$. One or more states $\phi_{h_\alpha}\in\mathcal H_{\text{topo}}$ can be identified as the ``primary states" in the following sense \emph{in the thermodynamic limit}:
\begin{enumerate}[label=\alph*).]
\item $\hat L_n\phi_{h_\alpha}\in \bar{\mathcal H}_{\text{topo}}$ for $n>0$
\item $\hat L_{-n}\phi_{h_\alpha}\in \mathcal H_{\text{topo}}$ for $n\ge 0$
\item $\hat C_{m,n}\phi_k\in \bar{\mathcal H}_{\text{topo}}$ for $m\neq n$
\end{enumerate}
In general $\hat L_n\phi_{h_\alpha}\neq 0$ for $n>0$, which violates the requirement for the primary fields in CFT. However, the assumption a). implies $\phi_{h_\alpha}$ are the highest weight states in $\mathcal H_{\text{topo}}$, thus qualifying them as the primary states in analogy to the primary fields. All descendant states that corresponds to the descendant fields in CFT are within $\mathcal H_{\text{topo}}$ from b). For the Virasoro algebra to hold, we need c), so the Virasoro algebra is satisfied within $\mathcal H_{\text{topo}}$. Thus the conformal invariance of $\mathcal H_{\text{topo}}$ is explicitly established.

We now look at the inner products and the norms of the states given by $\phi_k$. Let us use $|\phi_k\rangle$ to denote $\phi_k$, or the full wavefunction $\psi_k$ with the Gaussian factor. We define $\hat L_n|\phi_k\rangle$ as $\hat L_n$ acting on the holomorphic part of $\psi_k$. For the primary states we can define the inner product using the usual quantum mechanical overlap as follows:
\begin{eqnarray}\label{overlap}
\langle\phi_{h_\alpha}|\phi_{h_\beta}\rangle=\int dz_1dz_1^*\cdots dz_{N_e}dz_{N_e}^*\psi_{h_\alpha}^*\psi_{h_\beta}
\end{eqnarray}
A descendant state at level $N$ is thus given by $|\phi_{h_\alpha}^{\left(N\right)}\rangle = \hat L_{-k_1}\hat L_{-k_2}\cdots \hat L_{-k_n}|\phi_{h_\alpha}\rangle$, with $N=\sum_{i=1}^n k_i$. If the total angular momentum of $|\phi_{h_\alpha}\rangle$ is $M_\alpha$, then the total angular momentum of $|\phi_{h_\alpha}^{\left(N\right)}\rangle$ is $M_\alpha+N$. We can now define the following norm:
\begin{eqnarray}\label{norm}
&&\langle\phi_{h_\alpha}^{\left(N\right)}|\phi_{h_\alpha}^{\left(N\right)}\rangle\nonumber\\
&&=\langle\phi_{h_\alpha}|\hat L_{k_n}\hat L_{k_{n-1}}\cdots\hat L_{n_1}\hat L_{-k_1}\hat L_{-k_2}\cdots \hat L_{-k_n}|\phi_{h_\alpha}\rangle\quad
\end{eqnarray}
This is not equivalent to the quantum mechanical norm of $|\phi_{h_\alpha}^{\left(N\right)}\rangle$, since from Eq.(\ref{vv1}) and Eq.(\ref{vv2}) we can see that $\left(\hat L_n\right)^\dagger\neq \hat L_{-n}$. However, Eq.(\ref{norm}) can be evaluated in a well-defined way by commuting all of the $\hat L_{n}$ with $n\ge 0$ to the right using Eq.(\ref{vv3}), so that $\hat L_{k_n}\hat L_{k_{n-1}}\cdots\hat L_{n_1}\hat L_{-k_1}\hat L_{-k_2}\cdots \hat L_{-k_n}|\phi_{h_\alpha}\rangle$ is proportional to $|\phi_{h_\alpha}\rangle$. This is followed by the usual quantum mechanical overlap using Eq.(\ref{overlap}).

We thus have two types of norm or overlap between two states in $\mathcal H_{\text{topo}}$. The first type is the usual quantum mechanical overlap from the integration over $z_i,z_i^*$, in the form of Eq.(\ref{overlap}) for any two states. The other type is the so called ``conformal norm" and ``conformal overlap", which is computed from the Virasoro algebra, the highest weight condition of the primary state, and the orthonormality of the primary states (equivalent to the quantum mechanical overlap, for the primary states only). This distinction is important, since the linear dependence of a set of states depends entirely on the definition of the overlaps. At each level or total angular momentum sector, a set of states can be linearly dependent with the quantum mechanical overlap, but linearly independent with the conformal overlap, or vice versa. This is mainly because the Gram matrix of the conformal overlap is not positive definite. On the other hand, all physical quantities in principle should be derived from the quantum mechanical overlap.

The emergence of the central charge can be seen from the microscopic wavefunctions as follows:
\begin{eqnarray}
\langle&\phi_k&|[\hat L_n,\hat L_{-n}]|\phi_k\rangle=\langle\phi_k|\hat C_{n,n}|\phi_k\rangle\nonumber\\
&=&\sum\limits_{k'=0}^{n-1}k'\left(k'-n\right)\langle\phi_k|c^\dagger_{k'}\hat c_{k'}|\phi_k\rangle
\end{eqnarray}
where $\langle\phi_k|c^\dagger_{k'}\hat c_{k'}|\phi_k\rangle$ gives the average occupation of electrons in the $k'^{\text{th}}$ orbital. Since $|\phi_k\rangle$ is a state with edge excitations, physically there is only density modulation near the edge. In the thermodynamic limit and for any finite value of $k'$, we have $\langle\phi_k|c^\dagger_{k'}\hat c_{k'}|\phi_k\rangle=\nu$, the filling factor of the FQH phase. We thus have:
\begin{eqnarray}
\lim_{N_e\rightarrow\infty}\langle&\phi_k&|[\hat L_n,\hat L_{-n}]|\phi_k\rangle=\frac{2\nu}{12}\left(n^3-n\right)
\end{eqnarray}
and we can identify the central charge $c=2\nu$. The importance of this result will be discussed in the later sections.

From the definitions of Eq.(\ref{vv1}) and Eq.(\ref{vv2}), it is obvious all states $|\phi_k\rangle$ are eigenstates of $\hat L_0$, with eigenvalues given by the total angular momentum, which is also the conformal dimension. Thus there is at least one primary state with conformal dimension that scales with $N_e^2$, and becomes infinity in the thermodynamic limit. The norm of the descendant states from this primary state thus is thus always positive, using the definition from Eq.(\ref{norm}). We will also discuss this potentially interesting point in the specific examples later. 

\section{The case of $\lambda=0$}\label{lambda0}

We will now first focus on the idealised case of $\hat H_{\text{topo}}$, for which we assume the null space $\mathcal H_{\text{topo}}$ is well-defined. It turns out in many cases for the FQH systems, we can find model local Hamiltonians with well-defined null space, which allows us to scale the energies of states in $\bar{\mathcal H}_{\text{topo}}$ to infinity for any \emph{finite} systems. Thus $\mathcal H_{\text{topo}}$ can be realised by physically relevant Hamiltonians. For all the analysis in the previous section to apply in the thermodynamic limit, we also need the Hamiltonian to have a finite energy gap in the thermodynamic limit. This is necessary, because otherwise we cannot send the energies of the states in $\bar{\mathcal H}_{\text{topo}}$ to infinity by rescaling the Hamiltonian.

\subsection{The Laughlin and Moore-Read phase}

Let us look at two familiar examples of the Laughlin phase and the Moore Read phase. The model Hamiltonian of the former is the $\hat V_1^{\text{2bdy}}$ Haldane pseudopotential, while for the latter is the three-body interaction Hamiltonian which we denote as $\hat V_3^{\text{3bdy}}$. In both cases, the CFT description of the null space (denoted by $\mathcal H_{\text{topo}}^{L}$ for the Laughlin phase, and $\mathcal H_{\text{topo}}^{M}$ for the Moore Read phase) is well known, with unitary CFT models. For the Laughlin phase, the $\mathcal H_{\text{topo}}^L$ can be mapped to the Hilbert space of the $U(1)$ chiral bosons; while for the Moore Read phase the $\mathcal H_{\text{topo}}^M$ can be mapped to the Hilbert space of the $U(1)$ chiral bosons with the additional Ising fermions\cite{read,rr}.

We will now look into the explicit conformal invariance of $\mathcal H_{\text{topo}}^L$. We can identify the Laughlin wavefunction $\phi_{h_L}=\prod_{i<j}\left(z_i-z_j\right)^q$ at filling factor $\nu=1/q$ as the primary state, with $\hat L_1|\phi_{h_L}\rangle=0$. While $\hat L_n|\phi_{h_L}\rangle$ does not vanish for $n>1$, it clearly lives entirely in $\bar{\mathcal H}_{\text{topo}}^L$, since $|\phi_{h_L}\rangle$ is the highest density state in $\mathcal H_{\text{topo}}^L$ with the minimal total angular momentum $M_L=q\left(N_e^2-N_e\right)/2$. The entire space of $\mathcal H_{\text{topo}}^L$, which are spanned by Jack polynomials with $\alpha=-2/\left(q-1\right)$ and admissible root configurations (i.e. no more than one electron for every $q$ consecutive orbitals)\cite{bernevig0}, can be generated by repeated applications of $\hat L_{-n}, n>0$ on $|\phi_{h_L}\rangle$, with the well-known Virasoro level counting of $1,1,2,3,5,7,\cdots$ corresponding to the level $N=0,1,2,3,4,5,\cdots$. We have also numerically verified assumption (c) for $|\phi_k\rangle\in\mathcal H_{\text{topo}}^L$ with finite size scaling .

In this particular description, the conformal dimension of $|\phi_{h_L}\rangle$, or the eigenvalue of $\hat L_0$, is $h_L=\frac{q}{2}N_e\left(N_e-1\right)$. The central charge is given by $c=2/q$. For $q\ge 3$ we have $c<1$, and this corresponds to a nonunitary CFT for any finite $N_e$. However, states with negative conformal norm (as computed from Eq.(\ref{norm})) can only occur at very large angular momenta, at which the finite size effect comes in and the Virasoro counting is no longer obeyed. Thus from the quantum mechanical point of view, those states with negative conformal norm are just linear combinations of other states in the same angular momentum sector. In the thermodynamic limit when $h_L\rightarrow\infty$, the Virasoro counting is obeyed at any arbitrarily large angular momentum sector,and  all states will have positive conformal norm. It is thus in every sense a valid CFT description of $\mathcal H_{\text{topo}}^L$ with $h_L=\infty,c=2/q$, even though it is apparently quite different from the usual CFT description with chiral free bosons ($h=q,c=1$).

\begin{figure}
\includegraphics[width=\linewidth]{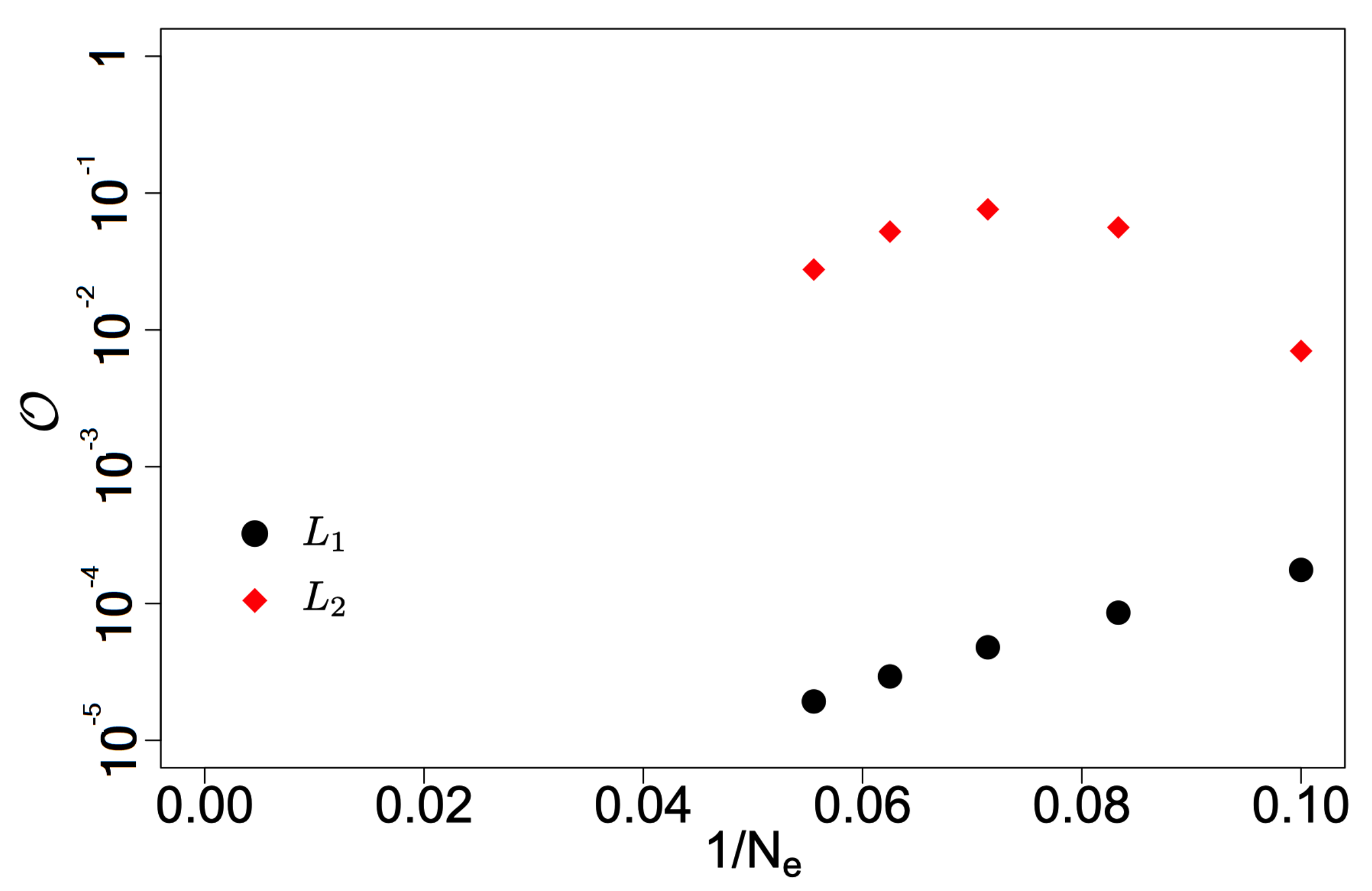}
\caption{The overlap with $\mathcal H_{\text{topo}}$ after appying $L_{1}$ (black plot) and $L_{2}$ (red plot) to the primary state $|\phi_{h'_M}\rangle$. The y-axis is the log of the overlap, and the x-axis is the inverse of the system size.}
\label{plot1}
\end{figure}

To extend this description to the Moore-Read phase with $\mathcal H_{\text{topo}}^M$, we can also identify the Pfaffian ground state $|\phi_{h_M}\rangle$ as the primary state with $h_M=N_e\left(N_e-\frac{3}{2}\right), c=1$. However, its conformal family does not span the entire $\mathcal H_{\text{topo}}^M$, which is the space of Jack polynomials with $\alpha=-3$ and admissible root configurations satisfying no more than two electrons in every four consecutive orbitals. In particular, at level $N=2$, there are three linearly independent states in $\mathcal H_{\text{topo}}^M$, while only two descendant states from $|\phi_{h_M}\rangle$, namely $\hat L_{-2}|\phi_{h_M}\rangle, \hat L_{-1}\hat L_{-1}|\phi_{h_M}\rangle$. We can thus construct the state at $N=2$ that is orthogonal to $\hat L_{-2}|\phi_{h_M}\rangle, \hat L_{-1}\hat L_{-1}|\phi_{h_M}\rangle$ (using the \emph{quantum mechanical overlap}), and denote it as $|\phi_{h'_M}\rangle$. There is very strong numerical evidence that $|\phi_{h'_M}\rangle$ is annihilated by $\hat L_1,\hat L_2$ (and thus $\hat L_n$ with $n>0$ following the Virasoro algebra), so this state can be identified as a second primary state in $\mathcal H_t^M$ (see Fig.(\ref{plot1})).

We will thus have the following decomposition of $|\phi_{h'_M}\rangle$:
\begin{eqnarray}
|\phi_{h'_M}\rangle=|\phi_{h_M}\rangle\otimes|\tau_{M}\rangle
\end{eqnarray}
Since $|\phi_{h'_M}\rangle$ has conformal dimension $h_M+2$, and that $\lim_{N_e\rightarrow\infty}\langle\phi_{h_M}|[\hat L_n,\hat L_{-n}]|\phi_{h_M}\rangle=\lim_{N_e\rightarrow\infty}\langle\phi_{h'_M}|[\hat L_n,\hat L_{-n}]|\phi_{h'_M}\rangle$ (the electron density at the center of the disk is not affected by excitations, or density modulations at the disk boundary), we then have a second primary state $|\tau_M\rangle$ with conformal dimension $h=2$ and central charge $c=0$. The fact that it has zero central charge is quite interesting. We will show it is a reasonable result that warrants further investigations. 

The level counting of the primary state with $h=2,c=0$ can be computed explicitly, and listed in the third column of Table.\ref{t1}. If we convolute the level counting from $|\tau_M\rangle$ with that of $|\phi_{h_M}\rangle$, the total number of states at $N>5$ over-counts those in $\mathcal H_{\text{topo}}^M$, indicating that not all states from the direct product of the two conformal families are linearly independent with the \emph{quantum mechanical overlap} (see the fifth column of Table.\ref{t1}). This again comes from the fact that the quantum mechanical overlap and the conformal overlap are not equivalent. Thus in principle, this mismatch does not contradict the conformal invariance of $\mathcal H_{\text{topo}}^M$.

On the other hand, the important discovery here is that if we only count the number of unitary states at each level using the conformal overlap (states with positive conformal norms), its convolution with the Virasoro counting of $|\phi_{h_M}\rangle$ agrees \emph{exactly} with the counting of the Pfaffian quasiholes, or the Hilbert space of $\mathcal H_{\text{topo}}^M$, for all the system sizes we can check numerically (see the sixth and seventh column of Table.\ref{t1}). This is also reasonable, because we should be able to map the CFT description here to the well-known CFT models with $U(1)$ bosons and Ising fermions (the $\mathcal M\left(4,3\right)$ minimal model), which is unitary. We conjecture this observation can be generalised to the entire Read-Rezayi series, which we will discuss in more details elsewhere.
\begin{widetext}
\begin{center}
\begin{table}[h!]
\centering
\begin{tabular}{ |c|c|c|c|c|c|c|c|c|} 
 \hline
 $N$ & $\{|\phi_{h_I}\rangle\}$ & $\{|\tau_{I}\rangle\}$ &$\{|\bar{\tau}_{I}\rangle\}$ & $\{|\phi_{h_I}\rangle\otimes|\text{vac}\rangle,|\phi_{h_I}\rangle\otimes|\tau_{I}\rangle\}$ &$\{|\phi_{h_I}\rangle\otimes|\text{vac}\rangle,|\phi_{h_I}\rangle\otimes|\bar{\tau}_{I}\rangle\}$ & $\mathcal H_{\text{topo}}^M$ &$\mathcal H_{\text{topo}}^G$ & $\mathcal H_{\text{topo}}^H$ \\ 
 \hline
  $0$ &1& $0$ & 0 &1 &1&1&1&1\\ 
 \hline
   $1$ &1& $0$ & 0 &1 &1&1&1&1\\ 
 \hline
  $2$ &2& $1$ & 1&3 &3&3&3&3\\ 
 \hline
  $3$ &3& $1$ & 1 &5 &5&5&5&5\\ 
 \hline
  $4$ &5& $2$ & 2 &10 &10&10&10&10\\ 
 \hline
  $5$ & 7&$2$ & 2 &16 &16&16&16&16\\ 
 \hline
  $6$ & 11&$4$ & 3 &29 &28&28&29&29\\ 
 \hline
  $7$ & 15&$4$ & 3 &45 &43&43&45&45\\ 
 \hline
   $8$ &22& $7$ & 5 &75 &70&70&74&75\\ 
 \hline
  $9$ &30& $8$ & 5&115 &105&105&113&115\\ 
 \hline
   $10$ &42& $12$ & 7 &181 &161&161&176&180\\ 
 \hline
\end{tabular}
\caption{Level counting of the CFT model $|\psi_{h_I}\rangle$ with $h=\infty,c=2\nu$, and $|\tau\rangle$ with $h=2,c=0$. $\{\bar\tau\rangle\}$ denotes the collection of the primary and descendant states in the conformal family with positive conformal norms. The fifth column gives the upper bound of the counting. The sixth and seventh column have the identical counting.}
\label{t1}
\end{table}
\end{center}
\end{widetext}

\subsection{The Gaffnian and the Haffnian phase}

The Gaffnian and the Haffnian states are closely related to the Pfaffian states, as they are all the highest density zero energy states of the leading three-body pseudopotential interactions. Let us denote the null space that contains the Gaffnian and Haffnian quasiholes as $\mathcal H_{\text{topo}}^G$ and $\mathcal H_{\text{topo}}^H$ respectively. While the model Hamiltonian for the Pfaffian is $\hat H_{\text{mr}}=\hat V_3^{\text{3bdy}}$, that of the Gaffnian is $\hat H_{\text{gf}}=\hat V_3^{\text{3bdy}}+\hat V_5^{\text{3bdy}}$, and that of the Haffnian is $\hat H_{\text{hf}}=\hat V_3^{\text{3bdy}}+\hat V_5^{\text{3bdy}}+\hat V_6^{\text{3bdy}}$. There has been no rigorous claims on if $\hat H_{\text{gf}}$ and $\hat H_{\text{hf}}$ are gapped in the thermodynamic limit (and no rigorous statement can be made for $\hat H_{\text{mr}}$ for that matter), but even if they are gapless (as supported by a number of arguments), their null spaces are still well-defined for any system sizes. $\mathcal H_{\text{topo}}^G$ is spanned by Jack polynomials with $\alpha=-3/2$ and admissible root configurations satisfying no more than two electrons for any five consecutive orbitals. The Haffnian states (and their quasiholes) are no longer Jack polynomials, but they can be uniquely determined using the LEC formalism. There is thus still a one-to-one correspondence from $\mathcal H_{\text{topo}}^H$ and root configurations satisfying no more than two electrons for any six consecutive orbitals.

In both cases, the highest density Gaffnian state and Haffnian state can be identified as the primary state, which we denote as $|\phi_{h_G}\rangle$ and $|\phi_{h_H}\rangle$ respectively. There is also an additional primary state with $h=2,c=0$, just like the Moore Read case, so we will also denote with $|\tau_{G}\rangle$ and $|\tau_{H}\rangle$. There is, however, an important difference here. While $\mathcal H_{\text{topo}}^M$ only consists of descendant states of $|\tau_M\rangle$ with positive conformal norms, this is no longer the case for $\mathcal H_{\text{topo}}^G$ and $\mathcal H_{\text{topo}}^H$. The latter contain descendant states from $|\tau_{G}\rangle$ or $|\tau_{H}\rangle$ with both positive and negative conformal norms. In fact, based on extensive numerical evidence, we conjecture this is the case for the null space of all three-body interaction Hamiltonians of the following form (note there is no three-body pseudopotential $\hat V_4^{\text{3bdy}}$):
\begin{eqnarray}\label{projection}
\hat H_{\text{3bdy}}=\sum_{k=3}^{k_0}\hat V_k^{\text{3bdy}}
\end{eqnarray}
which has been extensively analysed in Simon et.al\cite{simon3}. In all these cases, the highest density zero energy state is the primary state with infinite conformal dimension in the thermodynamic limit and central charge $2\nu$, which we can collectively denote as $|\phi_{h_I}\rangle$. Its conformal family thus give the full Virasoro counting that are linearly independent quantum mechanically with positive conformal norm. There is also one additional primary state with $h=2,c=0$, which we collectively denote as $|\tau_{I}\rangle$. All states at level $N$ are given as follows:
\begin{eqnarray}
&&|\phi_{h_\alpha,I}^{\left(N\right)}\rangle=\hat L_{-k_1}\hat L_{-k_2}\cdots\hat L_{-k_n}|\phi_{h_I}\rangle\otimes|\text{vac}\rangle\label{c1}\\
&&|\phi_{h_\beta,I}^{\left(N\right)}\rangle=\hat L_{-k'_1}\hat L_{-k'_2}\cdots\hat L_{-k'_{n'}}|\phi_{h_I}\rangle\nonumber\\
&&\qquad\qquad\otimes\hat L_{-k''_1}\hat L_{-k''_2}\cdots\hat L_{-k''_{n''}}|\tau_I\rangle\label{c2}
\end{eqnarray}
where $N=\sum_{i=1}^nk_i=\sum_{i=1}^{n'}k'_i+\sum_{i=1}^{n''}k''_i+2$, $\alpha,\beta$ are indices of states at level N, and both $|\phi_{h_\alpha,I}^{\left(N\right)}\rangle,|\phi_{h_\beta,I}^{\left(N\right)}\rangle$ are microscopic wavefunctions that can be explicitly obtained (in the LLL they are of holomorphic variables $z_1,z_2,\cdots z_{N_e}$).

The conformal family of $|\tau_I\rangle$ has singular descendant states (descendant states with zero conformal norm). By removing those singular states, $|\phi_{h_\alpha,I}^{\left(N\right)}\rangle,|\phi_{h_\beta,I}^{\left(N\right)}\rangle$ are linearly independent with respect to the conformal overlaps. However, $|\phi_{h_\alpha,I}^{\left(N\right)}\rangle,|\phi_{h_\beta,I}^{\left(N\right)}\rangle$ are over-complete and linearly dependent with respect to the quantum mechanical overlap for finite $k_0$, so the counting of the linearly independent microscopic wavefunctions in the null spaces of Eq.(\ref{projection}) is bounded from above by the counting given by Eq.(\ref{c1}) and Eq.(\ref{c2}) (after the singular fields are removed). We have checked explicitly for $k_0\le 8$.

The conformal family of $|\tau_{I}\rangle$ also contains fields with negative conformal norms, because the associated CFT model ($h=2,c=0$) is non-unitary and irrational. However, if we denote $\{|\bar{\tau}_{I}\rangle\}$ as a subset of the conformal family that contains only states with positive conformal norm, then the remaining counting from Eq.(\ref{c1}) and Eq.(\ref{c2}) (after removing the singular and negative norm states) agrees with the counting of $\mathcal H_{\text{topo}}^M$, which is the null space of Eq.(\ref{projection}) with $k_0=3$. As we increase $k_0$ to $5,6$, etc. (corresponding to $\mathcal H_{\text{topo}}^G,\mathcal H_{\text{topo}}^H$, and so on), more and more states from Eq.(\ref{c2}) with negative conformal norm are needed to match the counting of the null space. We thus conjecture the full conformal counting (all negative conformal norm states) is needed to account for the counting of the null space in the limit of $k_0\rightarrow\infty$.

One of the most important aspects of the FQH edge dynamics is the thermal Hall conductance\cite{banerjee1,banerjee2,bernevig1}, normally related to the central charge of the conformal edge theory and should be completely determined by the null space $\mathcal H_{\text{topo}}$. It seems in the description described here, the central charge of $2\nu$ and $0$ comes from the bulk properties at the center of the quantum Hall droplet, thus unrelated to what happens at the edge. However, the thermal Hall conductance is completely determined by the ``density of states", or the counting of quasihole states in each angular momentum sector\cite{bernevig1}. For the Laughlin and Moore-Read state, the counting can be completely determined by the $h=\infty, c=2\nu$ and $h=2,c=0$ CFT models in the new description here. For ``non-unitary" states like the Gaffnian and Haffnian states, we are not sure at this stage how the non-unitary counting of the $h=2,c=0$ model is gradually incorporated, as $k_0$ in Eq.(\ref{projection}) increases. Yet this is also the case for the conventional CFT description with the non-unitary and/or irrational CFT models, as the central charge is different from the ``effective central charge" \cite{read2}obtained from the counting of $\mathcal H_{\text{topo}}^G$ and $\mathcal H_{\text{topo}}^H$. The latter can only be obtained microscopically (i.e. from the properties related to the Jack polynomials).

In this sense, there is no obvious disadvantage with the new CFT description presented here. While we do not have a rigorous proof available, we believe the description here can be mapped to the usual CFT descriptions with minimal or irrational models. The results also show the strong relevance of the $h=2,c=0$ CFT model for the three-body projection Hamiltonians, including the Moore-Read, Gaffnian and Haffnian states. It illustrates that the conformal invariance of the topological null space can be described from different perspectives with effective CFT models. Many physical features of the null space, which are related to the edge dynamics, however, depends on the quantum mechanical properties of the many-body wavefunctions that could be fundamentally different from the CFT descriptions with their own definitions of hermitian conjugates, norms and overlap, as well as quasihole correlations. After all, the Gaussian factors in the many-body wavefunctions play an important role for the quantum mechanical norm, overlaps and thus linear dependence of those wavefunctions. Such Gaussian factors are not accounted for in the CFT descriptions, and they explicitly break conformal invariance with the presence of the magnetic length. We thus need to be more careful in characterising the physical properties of the Gaffnian and Haffnian phase solely from the effective CFT descriptions.

\section{Gaffnian and Haffnian states as elementary excitations}\label{excitations}

To further motivate the physical relevance of the Gaffnian and Haffnian phases to the gapped FQH systems, we look at the familiar Laughlin phase at $\nu=1/3$, and focus on its gapped elementary excitations (namely the neutral and quasielectron excitations). Such excitations are well studied in the composite fermion picture and the Jack polynomial formalism\cite{yangbo1,sreejith,ajit1,ajit2,yangbo2,gromov}. The elementary low-lying neutral excitation is the magnetoroton mode. In the long wavelength limit, it is a quadrupole excitations from the geometric deformation of the ground state. At large momenta, it is a dipole excitation consisting of a pair of separated Laughlin quasielectron and quasihole. The energy gap of the magnetoroton mode defines the incompressibility, and thus the robustness of the Hall plateau, of the Laughlin phase.

From the microscopic point of view, the best way to understand the physical properties of such gapped excitations is to construct good model wavefunctions. Unlike the null space of model Hamiltonians (the zero energy ground states and quasiholes), in general microscopic wavefunctions for the gapped excitations are non-universal, and different approaches lead to (slightly) different model wavefunctions. The magnetoroton mode is special in the sense that the composite fermion approach and the Jack polynomial formalism give exactly identical model wavefunctions. This is true for excitations consisting of only one Laughlin quasielectron (thus including single quasielectron states that are also gapped). For states containing more than one quasielectrons, the two approaches yields different model wavefunctions, though their overlaps are generally quite high.

Given the uniqueness of the magnetoroton model wavefunctions, we show here that the quadrupole excitations are exact zero energy states of the Haffnian model Hamiltonians, thus are states within $\mathcal H_{\text{topo}}^H$. In contrast, the dipole excitations, as well as the single Laughlin quasielectron states, are exact zero energy states of the Gaffnian model Hamiltonian, thus living within $\mathcal H_{\text{topo}}^G$\cite{yangbo3,yangbo4}. To see that, let us write the root configuration of the magnetoroton modes as follows\cite{yangbo1}:
\begin{eqnarray}\label{root1}
&&\d{1}\d{1}0\textsubring{0}\textsubring{0}0100100100100\cdots\qquad\Delta M=-2\quad\in\mathcal H_{\text{topo}}^H\quad \\
&&\d{1}\d{1}0\textsubring{0}010\textsubring{0}0100100100\cdots\qquad\Delta M=-3\quad\in\mathcal H_{\text{topo}}^G\quad \\
&&\d{1}\d{1}0\textsubring{0}010010\textsubring{0}0100100\cdots\qquad\Delta M=-4\quad\in\mathcal H_{\text{topo}}^G\quad \\
&&\d{1}\d{1}0\textsubring{0}010010010\textsubring{0}0100\cdots\qquad\Delta M=-5\quad\in\mathcal H_{\text{topo}}^G\quad \\
&&\qquad\qquad\vdots\nonumber
\end{eqnarray}
Here the solid and open circles beneath the digits indicate the locations of quasiparticles (of charge $e/3$, when three consecutive orbitals contain more than one electron) and quasiholes (of charge $-e/3$, when three consecutive orbitals contain fewer than one electron). Each root configuration represents a many-body wavefunction, where only basis ``squeezed" from the root configuration have non-zero coefficients. These non-zero coefficients can also be uniquely determined using the method in\cite{yangbo1}, from which the model wavefunctions are obtained.

The easiest way to see that the $\Delta M=-2$ state is a zero energy state of the Haffnian model Hamiltonian is that it satisfies the local exclusion condition (LEC)\cite{yangbo5,yangbo6} of $\{4,2,4\}$ at the center of the disk (corresponding to the north pole of the sphere). Similarly, we know the $\Delta M<-2$ states are the zero energy states of the Gaffnian model Hamiltonian, because they satisfy the LEC of $\{2,1,2\}\lor\{5,2,5\}$. Note that near the filling factor of $\nu=1/3$, there is an extensive number of Haffnian or Gaffnian quasiholes.

Let us denote the subspace of Laughlin ground state and quasiholes (the null space of $\hat V_1^{\text{2bdy}}$ pseudopotential) to be $\mathcal H_{\text{topo}}^L$, then we clearly have the relationship that $\mathcal H_{\text{topo}}^L\in\mathcal H_{\text{topo}}^H\in\mathcal H_{\text{topo}}^G$. We can thus reinterpret the elementary neutral excitations of the Laughlin phase as the quantum fluids of interacting Haffnian or Gaffnian quasiholes\cite{yangbo4}. Similarly, if we look at the model wavefunctions of a single Laughlin quasielectron, it has the following root configuration:
\begin{eqnarray}
\d{1}\d{1}0\textsubring{0}0100100100100\cdots\qquad\Delta M=-N_e/2\quad\in\mathcal H_{\text{topo}}^G\qquad
\end{eqnarray}
It is also a zero energy state of the Gaffnian model Hamiltonian. For multiple quasielectrons that are far away from each other, they can all be considered as some locally bound states of Gaffnian quasiholes.

\subsection{The Gaplessness of the Model Haffnian Hamiltonian}

The model Hamiltonian of the Haffnian state is the special case of Eq.(\ref{projection}) with $k_0=6$. It in fact is given by a family of the following Hamiltonian:
\begin{eqnarray}\label{haffnian}
\hat H_{\text{hf}}=\hat V_3^{\text{3bdy}}+\lambda_1\hat V_5^{\text{3bdy}}+\lambda_2\hat V_6^{\text{3bdy}}
\end{eqnarray}
with $\lambda_1,\lambda_2>0$. The null space of Eq.(\ref{haffnian}) is $\mathcal H_t^H$. One should note that if we define the the Laughlin ground states and quasiholes space (i.e. the null space of $\hat V_1^{\text{2bdy}}$) as $\mathcal H_{\text{topo}}^L$, we then have $\mathcal H_{\text{topo}}^L\in\mathcal H_{\text{topo}}^H$.

What we are able to show here, is that the incompressibility of $\hat V_1^{\text{2bdy}}$ at $\nu=1/3$ (more specifically at $N_o=3N_e-2$, where $N_o$ is the number of orbitals on the sphere or disk geometry) implies that Eq.(\ref{haffnian}) is gapless  at $N_o=3N_e-4$ (where the Haffnian state is the highest density zero energy state), for any positive values of $\lambda_1,\lambda_2$. This is because Laughlin quasielectrons, which are orthogonal to the Haffnian ground state in the thermodynamic limit, do not have a finite energy gap with Eq.(\ref{haffnian}).

The incompressibility of $\hat V_1^{\text{2bdy}}$ implies both the quadrupole and dipole excitations cost a finite amount of energy in the thermodynamic limit. We now look at a two dimensional subspace spanned by two states of the following root configuration:
\begin{eqnarray}
&&110000110000110000\cdots 11000011\label{h1}\\
&&11000100100100\cdots 100100100011\label{h2}
\end{eqnarray}
The first state of Eq.(\ref{h1}) is the Haffnian model state, which is the unique zero energy state in $L=0$ with $N_o=3N_e-4$ from Eq.(\ref{haffnian}). The second state of Eq.(\ref{h2}) is the Laughlin state with two quasielectrons in the same quantum sector. It can be constructed using the method in\cite{yangbo4}. The model state has very high overlap with the exact ground state of $\hat V_1^{\text{2bdy}}$ in $L=0$ with $N_o=3N_e-4$ (the arguments here also applies if we use this exact ground state in place of Eq.(\ref{h2})). Since the Haffnian state contains an extensive number of quadrupole excitations, its variational energy with respect to $\hat V_1^{\text{2bdy}}$ is also extensive. On the other hand, the variational energy of the two-quasielectron state of Eq.(\ref{h2}) is finite in the thermodynamic limit (see Fig.(\ref{plot2}b)), which is double the Laughlin charge gap at $\nu=1/3$.

We now look at the spectrum of Eq.(\ref{haffnian}) within this two dimensional subspace and argue that the two energies have to be \emph{degenerate} in the thermodynamic limit. The overlap of the two states Eq.(\ref{h1}), Eq.(\ref{h2}) quickly decays with the system size (see Fig.(\ref{plot2}c)), so for all purposes we can treat them as the eigenstates in this subspace, with the Haffnian state as the zero energy ground state. If Eq.(\ref{h2}) has a finite energy gap in the thermodynamic limit, then we can consider perturbing Eq.(\ref{haffnian}) with an infinitesimal amount of $\hat V_1^{\text{2bdy}}$ as follows:
\begin{eqnarray}
\hat H=\hat H_{\text{hf}}+\lambda\hat V_1^{\text{2bdy}}
\end{eqnarray}
there will be a level crossing no matter how small $\lambda$ is, since the Haffnian state will have infinite energy in the thermodynamic limit, while Eq.(\ref{h2}) will have finite energy. This is not possible unless Eq.(\ref{h1}) and Eq.(\ref{h2}) are degenerate at $\lambda=0$, implying that Eq.(\ref{haffnian}) is gapless in the thermodynamic limit. In another word, if $\hat H_{\text{hf}}$ is gapped in the thermodynamic limit, then an infinitesimally small perturbation can close the gap and lead to level crossing.
\begin{figure}
\includegraphics[width=9cm]{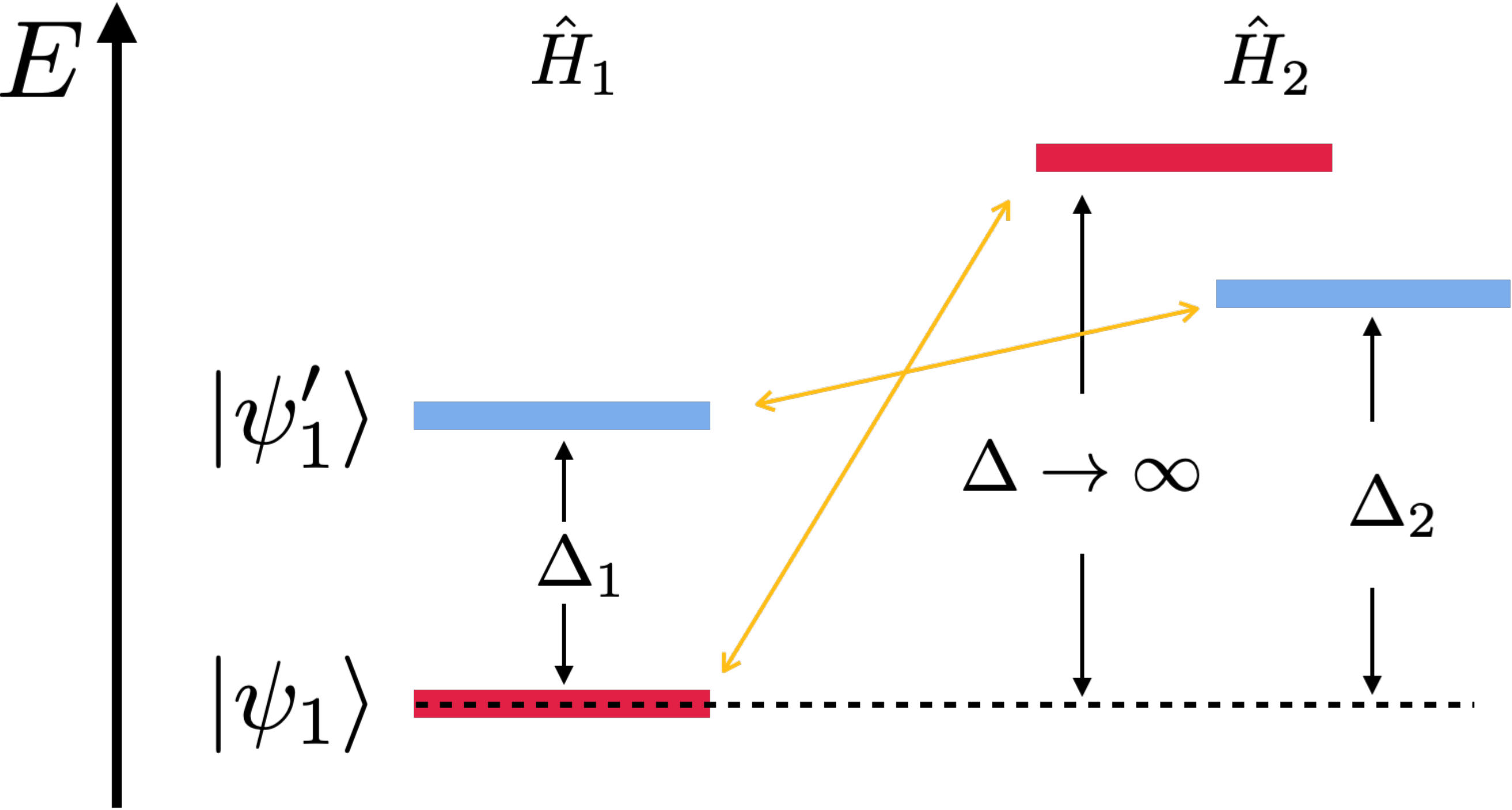}
\caption{We can set the energy of $|\psi_1\rangle$ with respect to $\hat H_1$ to be zero, so that of $|\psi_1'\rangle$ is $\Delta_1$. The variational energy of $|\psi_1'\rangle$ with respect to $\hat H_2$ is $\Delta_2$, while that of $|\psi_1\rangle$ is $\Delta$, which goes to infinity in the thermodynamic limit. An infinitesimally small perturbation of $\hat H_2$ to $\hat H_1$ will lead to a level crossing between the two states indicated by the yellow arrow.}
\label{fig4}
\end{figure} 

The argument can be generalised as follows. Let $\hat H_1,\hat H_2$ to be two local Hamiltonians with the null spaces $\mathcal H_1,\mathcal H_2$ respectively. Let $|\psi_1\rangle\in\mathcal H_1,|\psi_2\rangle\in\mathcal H_2$ be the highest density states (with densities $\rho_1\ge\rho_2$) in their respective null spaces. If there exists a state $|\psi_1'\rangle$ with density $\rho_1'$ and finite values $\Delta_1\ge 0,\Delta_2\ge 0$ such that following conditions are satisfied:
\begin{eqnarray}
&&\qquad\rho_1'\ge\rho_1\label{c1}\\
&&\lim\limits_{N_e\rightarrow\infty}\langle\psi'_1|\hat H_1|\psi'_1\rangle=\Delta_1\label{c4}\\
&&\lim\limits_{N_e\rightarrow\infty}\langle\psi_1|\hat H_2|\psi_1\rangle\rightarrow\infty\label{c2}\\
&&\lim\limits_{N_e\rightarrow\infty}\langle\psi'_1|\hat H_2|\psi'_1\rangle=\Delta_2\label{c3}
\end{eqnarray}
then we have 
\begin{eqnarray}
\lim\limits_{N_e\rightarrow\infty}\langle\psi_1|\hat H_1|\psi_1\rangle=\lim\limits_{N_e\rightarrow\infty}\langle\psi'_1|\hat H_1|\psi'_1\rangle
\end{eqnarray}
implying $\hat H_1$ is gapless with neutral excitations (if $\rho_1=\rho_1'$) or charged excitations (if $\rho_1<\rho'_1$). Note that $|\psi_1\rangle$ and $|\psi_1'\rangle$ are two eigenstates of $\hat H_1$. If their energies are gapped in the thermodynamic limit, an infinitesimally small amount of perturbation of $\hat H_2$ to $\hat H_1$ will close the gap if Eq.(\ref{c1}) - Eq.(\ref{c3}) are satisfied. A simple schematic illustration of the argument can be found in Fig.(\ref{fig4}).

We do not require $\hat H_2$ to be gapped in the above arguments. However in the case of $\hat H_1=\hat H_{\text{hf}},\hat H_2=\hat V_1^{\text{2bdy}}$, we have $\rho_1>\rho_2$ but $\lim\limits_{N_e\rightarrow\infty}\left(\rho_2-\rho_1\right)= 0$. Thus we can find the states $|\psi'_1\rangle$ with $\rho'_1\ge\rho_1$ such that they are the Laughlin ground state plus a finite number of Laughlin quasielectrons, which are exponentially localised excitations. These states satisfy Eq.(\ref{c1}), Eq.(\ref{c4}) and Eq.(\ref{c3}) (note the Laughlin ground state is also in the null space of $\hat H_{\text{hf}}$). From the fact that $\hat V_1^{\text{2bdy}}$ is gapped, we know its quadrupole excitation is gapped in the thermodynamic limit. Given $|\psi_1\rangle$, or the Haffnian state, contains an extensive number of quadrupole excitations, Eq.(\ref{c2}) is also satisfied. Thus $\hat H_{\text{hf}}$ has both gapless neutral and charged excitations and is compressible. All states that physically represents the Laughlin ground state with a finite number of quasielectrons are degenerate with the Haffnian ground state in the thermodynamic limit if the interaction Hamiltonian is $\hat H_{\text{hf}}$.

\begin{figure}
\includegraphics[width=\linewidth]{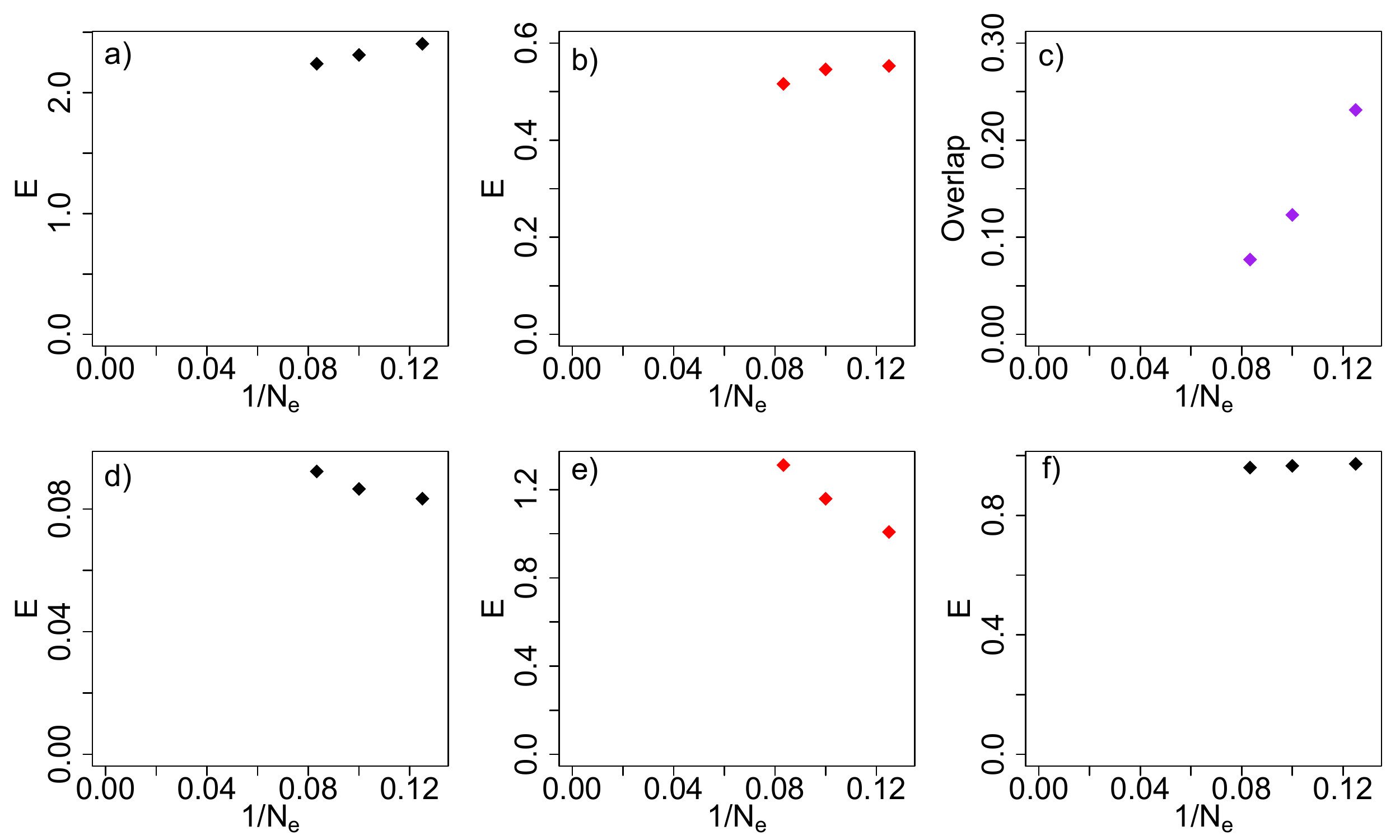}
\caption{The x-axis is the inverse of system size (number of electrons), and the y-axis are a).Variational energy of the two-quasielectron Laughlin state with respect to $\hat H_{\text{hf}}$; b).Variational energy of the two-quasielectron Laughlin state with respect to $\hat V_1^{\text{2bdy}}$;c). Overlap of the two-quasielectron Laughlin state and the Haffnian model state; d).Variational energy of the $\hat V_{\text{LLL}}$ ground state at $\nu=2/5$ with respect to $\hat H_{\text{gf}}$; e). Variational energy of the $\hat V_{\text{LLL}}$ ground state at $\nu=2/5$ with respect to $\hat V_1^{\text{2bdy}}$; f). Overlap between the $\hat V_{\text{LLL}}$ ground state at $\nu=2/5$ and the Gaffnian state. }
\label{plot2}
\end{figure} 

\subsection{The Gaffnian state at $\nu=2/5$}

\begin{figure}
\text{             {\color{white} jkjkj}}
\includegraphics[width=9cm]{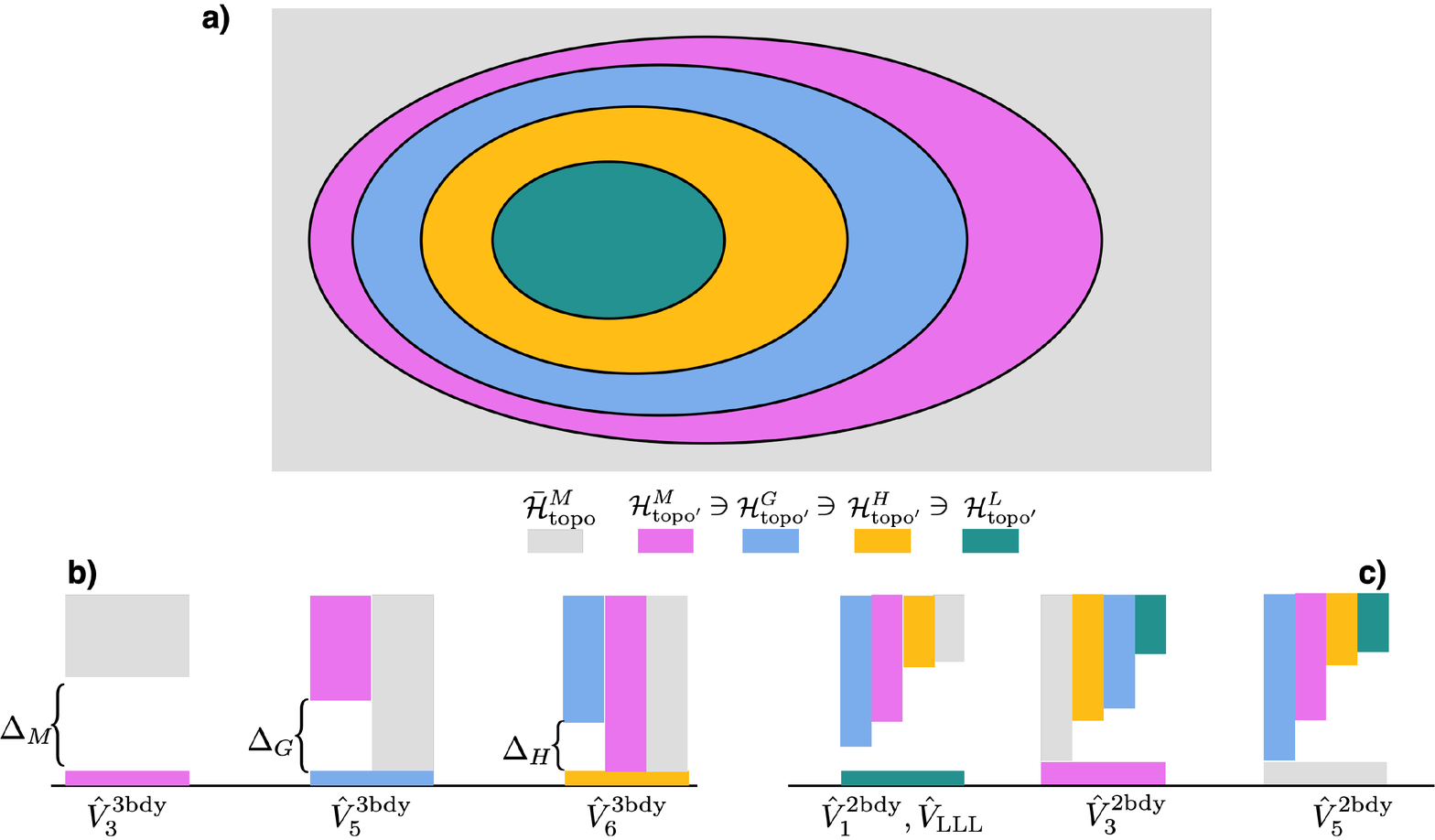}
\caption{a). The relationship between different Hilbert spaces defined in the main text. b). The variational energies of different Hilbert spaces with three-body pseudopotentials. c). The variational energies of different Hilbert spaces with two-body interactions. The variational energies are computed numerically from finite size systems. While the relative strength of different Hilbert spaces is consistent for different system sizes, they should be just indicative of what could happen in the thermodynamic limit.}
\label{fig2}
\end{figure}

The argument above does not apply to the Gaffnian state, because there is no known local Hamiltonians with well-defined null spaces nearby $\hat H_{\text{gf}}$, playing the role of $\hat V_1^{\text{2bdy}}$ to $\hat H_{\text{hf}}$. From the three-body interactions, we know that $\bar{\mathcal H}_{\text{topo}}^M$ contains states with clusters of three particles having total relative angular momentum $3$. The Hilbert space of $\mathcal H_{\text{topo}}^M$\textbackslash$\mathcal H_{\text{topo}}^G$ contains states with clusters of three particles having total relative angular momentum $5$. The Hilbert space of $\mathcal H_{\text{topo}}^G$\textbackslash$\mathcal H_{\text{topo}}^H$ contains states with clusters of three particles having total relative angular momentum $6$. These subspaces will thus be affected by individual three-body pseudopotentials differently as shown in Fig.(\ref{fig2}). In Fig.(\ref{fig2}), it is generally believed that $\lim\limits_{N_e\rightarrow\infty}\Delta_M$ is finite, and we have argued that $\lim\limits_{N_e\rightarrow\infty}\Delta_H=0$, with the gap closing from excitations in the subspace of $\mathcal H_{\text{topo}}^G$\textbackslash$\mathcal H_{\text{topo}}^H$ (though there could be other gapless modes). Thus for $\hat H_{\text{gf}}$ to be gapless, with positive $\hat V_3^{\text{3bdy}}$ the gapless mode can only be in the subspace of $\mathcal H_{\text{topo}}^M$\textbackslash$\mathcal H_{\text{topo}}^G$, which in particular are the zero energy states of $\hat V_3^{\text{3bdy}}$. One competing state for the Gaffnian state is the Abelian Jain state at the same filling factor and topological shift, with the following root configuration:
\begin{eqnarray}\label{jain}
11001001010010100100\cdots 10010011
\end{eqnarray}
which is unsqueezed from the root configuration of the Gaffnian state. The Jain state \emph{cannot} be uniquely determined by any known local operators, and there are several highest weight states\cite{regnault} supported by the basis squeezed by Eq.(\ref{jain}) (in contrast there is a unique highest weight state supported by the basis squeezed from the Gaffnian root configuration). For finite systems, the Jain state has very high overlap with the Gaffnian state, suggesting the basis \emph{unsqueezed} from the Gaffnian root configuration but \emph{squeezed} from Eq.(\ref{jain}) play a minor role. The Jain root configuration also shows that the Jain state contains two Gaffnian neutral excitations (one at the north pole, the other at the south pole).

Let us denote the Jain state as $|\psi_{J,2/5}\rangle$. From the root configuration, the basis of the Jain state clearly satisfies the LEC condition of $\{3,2,3\}$, it is thus a zero energy state of $\hat V_3^{\text{3bdy}}$, i.e. $|\psi_{J,2/5}\rangle\in\mathcal H_{\text{topo}}^M$. It is generally believed at the critical point\cite{simon2,jolicoeur,freedman} of $\hat H_{\text{gf}}$, $|\psi_{J,2/5}\rangle$ and $|\psi_{\rho_{\text{max}}}^G\rangle$ are degenerate in the thermodynamic limit (the latter is the Gaffnian state), based on the CFT conjecture. It is thus natural to consider the possibility of letting $|\psi_{J,2/5}\rangle$ play the role of $|\psi_1'\rangle$, with $|\psi_{\rho_{\text{max}}}^G\rangle=|\psi_1\rangle$ in Eq.(\ref{c1}) to Eq.(\ref{c3}). Even though there is no known model Hamiltonian for the Jain state, we note that the arguments above do not require $\hat H_2$ to be a local Hamiltonian. We can thus assume the existence of $\hat H_2$ such that Eq.(\ref{c3}) is satisfied.

However, it is not clear if such $\hat H_2$ can satisfy Eq.(\ref{c2}), given that $|\psi_{J,2/5}\rangle$ and $|\psi_{\rho_{\text{max}}}^G\rangle$ have very high overlap for finite systems (comparable to the overlap between the Laughlin model state and the LLL Coulomb interaction ground state). A more serious issue is the strong numerical evidence against Eq.(\ref{c4}). From Fig.(\ref{plot2}d) we see that the variational energy of the ground state with the \emph{lowest Landau level Coulomb interaction} with the Gaffnian model Hamiltonian seems to be extensive. This is qualitatively the same if the exact Jain $2/5$ state is used\cite{ajitunpublished}, i.e. $\langle\psi_{J,2/5}|\hat H_{\text{gf}}|\psi_{J,2/5}\rangle\sim O\left(N_e\right)$. While we can only access relatively small system sizes here, the numerics does suggest that as far as the ground state properties are concerned, the Gaffnian state and the Jain state seem to be indistinguishable topologically. We will discuss about the universal properties of their respective quasihole states in the next section.

\section{The case of $\lambda>0$}\label{realistic}

We now move on to more realistic interactions with $\lambda>0$ in Eq.(\ref{master}). Several possible scenarios can happen as we outlined in Sec.~\ref{intro}, and we look at these possibilities with the particular focuses at filling factor $\nu=2/5$ (where the Gaffnian state is located) and $\nu=1/3$ (where the Haffnian state is located). When we move away from the model Hamiltonian, the conformal invariance of the null spaces is broken, so in principle the connection to the CFT models is no longer valid. The two well-known FQH phases at these two filling factors are the Abelian Jain state and the Laughlin state. We would like to show, however, the physical relevance of the null spaces (i.e. quasihole subspaces) at these two filling factors. In particular, the goal is to see if the observed experimental data can be explained using their respective quasihole subspaces, and what new experimental results we can predict from the perspective of quasihole subspaces.

\subsection{The Gaffnian and the Jain phase}

One important question to ask is if the Gaffnian state and the Jain ground state at $\nu=2/5$ are topologically equivalent: that any topological indices computed from these two microscopic wavefunctions are identical\cite{yang1}. If $\hat H_G$ is gapped in the thermodynamic limit in the $L=0$ sector, then the statement has to be true even if the gap closes in some other $L$ sector. This is because the two states are adiabatically connected, and without gap closing in $L=0$ sector any physical properties computed from the two states have to go smoothly from one value to another, implying the invariance of topological indices.

We now assume $\hat H_G$ is gapless in the thermodynamic limit in the $L=0$ sector. Since we know $\hat V_3^{\text{3bdy}}$ gaps out $\bar{\mathcal H}_{\text{topo}}^M$, the gap can only close within $\mathcal H_{\text{topo}}^M$, which is where the Jain trial wavefunction from the composite fermion construction is located. Let us define $\mathcal H_{\text{topo}'}^M=\mathcal H^M_{\text{topo}}$\textbackslash$\mathcal H_{\text{topo}}^G$, then the Gaffnian ground state $|\psi_{0,G}\rangle$ is orthogonal to $\mathcal H_{\text{topo}'}^M$. The main question here is which subspace describes the low energy physics as $\lambda$ increases from zero. Let $\{|\psi_M\rangle\}\in\mathcal H_{\text{topo}}^M$ be the set of states degenerate with $|\psi_{0,G}\rangle$ in the thermodynamic limit. If we perturb $\hat H_G$ with an infinitesimal amount of $\hat V_1^{\text{2bdy}}$, we can apply the degenerate perturbation theory to the first order, and diagonalise $\hat V_1^{\text{2bdy}}$ in the subspace of $\{|\psi_M\rangle\}\cup |\psi_{0,G}\rangle$. Since $\{|\psi_M\rangle\}$ is orthogonal to $|\psi_{0,G}\rangle$, we expect the diagonal matrix elements to be extensive (thus going to infinity), while matrix elements between $\{|\psi_M\rangle\}$ and $|\psi_{0,G}\rangle$ to vanish, in the limit of large system sizes. From Fig.(\ref{fig2}) we expect $\langle\psi_{0,G}|\hat V_1^{\text{2bdy}}|\psi_{0,G}\rangle<E_M$, where $E_M$ is the ground state of $\hat V_1^{\text{2bdy}}$ within $\mathcal H_{\text{topo}'}^M$. We expect this to be true in the thermodynamic limit. Thus with the following Hamiltonian:
\begin{eqnarray}\label{masterg}
\hat H=\left(1-\lambda\right)\hat H_G+\lambda\hat V_1^{\text{2bdy}}
\end{eqnarray}
there is no level crossing between $|\psi_{0,G}\rangle$ and $\mathcal H_{\text{topo}'}^M$ when we increase $\lambda$ from $0$. Note that at $\lambda=0$, even if the Hamiltonian is gapless, the variational energies of $\hat H_G$ in $\mathcal H_{\text{topo}'}^M$ can only approach zero asymptotically. Since it is generally believed that the ground state of Eq.(\ref{masterg}) with $\lambda>0$ is adiabatically connected to the Abelian Jain state\cite{jolicoeur}, we argue that the Gaffnian \emph{ground state} is indeed adiabatically connected, and thus topologically equivalent, to the Jain \emph{ground state} from the composite fermion construction.

\subsubsection{The thermal Hall effect and the quasihole bandwidth}

The arguments above suggest that as far as the ground states are concerned, the Gaffnian phase and the Jain phase are topologically equivalent. The two phases, however, are \emph{not} topologically equivalent with regard to the universal properties of the low-lying excitations, which in particular dictates the (non-)Abelian-ness of the FQH phase. For a dilute gas of Gaffnian quasiholes, we also expect no level crossing between $\mathcal H_{\text{topo}}^G$ and $\mathcal H_{\text{topo}'}^M$ as $\lambda$ increases. We thus conjecture \textbf{P1}(b) should capture the adiabatic tuning from the Gaffnian model Hamiltonian to the realistic short-range interaction that attributes to the Hall plateau observed at $\nu=2/5$. We do not expect to see non-Abelian braiding of the quasiholes, because with realistic interaction a large bandwidth of the quasihole manifold will develop, lifting the required degeneracy\cite{toke}.

This effect should also be reflected in thermal quantum Hall measurement\cite{banerjee1,banerjee2}, which is given by the heat capacity of the chiral edge at the boundary of the quantum Hall fluid\cite{bernevig1}. Let the partition function of the $1D$ edge system be given as follows:
\begin{eqnarray}
\mathcal Z=\sum\limits_{N=0}^\infty g\left(N,\beta\right)e^{-\beta\epsilon_N}
\end{eqnarray}
where $\beta=1/k_BT$, and $g\left(N,0\right)$ is the number of the quasihole states at level N, i.e. the density of state as a function of the angular momentum. The ``kinetic energy" $\epsilon_N$ accounts for the energy of the states at the same level, which can be contributed from the confining potential of the quantum Hall droplet near the edge. Let $N$ have the unit of angular momentum, for general cases we can expand it as follows:
\begin{eqnarray}
\epsilon_N=c_1N+c_2N^2+\cdots
\end{eqnarray}
where $c_1=\frac{v_F}{2\pi L}$, with $v_F$ the fermi velocity and $L$ the circumference of the droplet. At finite temperature we have:
\begin{eqnarray}
g\left(N,\beta\right)=\sum\limits_\alpha e^{-\beta\epsilon_{N,\alpha}}
\end{eqnarray}
where $\alpha$ is the index of the states in a single level, and $\epsilon_{N,\alpha}$ is the additional energy costs from the creation energies of quasiholes as well as interaction between quasiholes. We can also absorb the non-linearity of the kinetic energy into the density of the states part, and rewrite the partition function as follows:
\begin{eqnarray}
&&\mathcal Z=\sum\limits_{N=0}^\infty \tilde g\left(N,\beta,c_2\right)e^{-\beta\epsilon_N}\label{partition}\\
&&\tilde g\left(N,\beta,c_2\right)=\sum\limits_\alpha e^{-\beta\left(\epsilon_{N,\alpha}+c_2N^2+\cdots\right)}
\end{eqnarray}
Conformal invariance implies $\tilde\epsilon_{N,\alpha}=\epsilon_{N,\alpha}+c_2N^2+\cdots=0$ identically. With this assumption, the thermal Hall conductance is given by $\kappa=\frac{c\pi^2k_B^2T}{3h}$, where $c$ is the central charge of chiral Luttinger liquid. In the composite fermion picture, the Jain state consists of two occupied CF levels analogous to the IQHE with $\nu=2$, so the central charge is $c=2$. One should note that even with the assumption of conformal invariance, the value of $c=2$ is not yet supported by the microscopic CF theory. This is because the quasihole excitations from each CF level with the CF construction are not orthonormal with each other, and there are missing states after the projection into the LLL\cite{yang1,ajit3}. These missing states will effectively reduce the density of states at each momentum sector. The effective field theories from the CF construction, on the other hand, are generally formulated from the CF theory before LLL projection, ignoring the missing states. Thus strictly speaking the effective field theories from the CF construction predicts an upper-bound of $c=2$ for the thermal Hall conductance. If we do not consider the non-universal factors in actual experiments that break the chiral Luttinger liquid description of the quantum Hall edge, the actual central charge from the thermal Hall conductance measurement should also be bounded above by $c=2$.

For the Gaffnian model Hamiltonian, the null space has an effective central charge $c=1+3/5$ from its Virasoro counting, which can be computed from Eq.(\ref{partition}) by taking $\tilde g\left(N,\beta,c_2\right)=\tilde g\left(N,0,0\right)$. It is important to note this effective central charge is different from the negative central charge predicted from the $\mathcal M\left(5,3\right)$ minimal model conventionally associated with the Gaffnian phase\cite{read2,bernevig1}. The negative central charge (thus the thermal Hall conductance) comes from the unphysical negative conformal norms. Such contributions have to be corrected, since all physical quasihole states have positive quantum mechanical norm and will contribute positively to the thermal Hall conductance.

With the realistic interaction and confining potential, conformal invariance is explicitly broken. This is reflected by $\tilde\epsilon_{N,\alpha}\neq 0$, which effectively modifies the density of state $\tilde g\left(N,\beta,c_2\right)$. From this perspective, the thermal Hall conductance is only universal in the presence of conformal symmetry, when it is independent of the fermi velocity $v_F$. We thus believe while the composite fermion description is a good effective theory at $\nu=2/5$, the thermal Hall conductance will not in general be quantized with $c=2$. It can be computed by assuming all edge excitations are Gaffnian quasiholes. The actual value, however, will depend strongly on the realistic interaction, which is known to split the quasihole bands due to the non-zero creation energies of quasiholes\cite{toke}.

Experimentally, the $\nu=2/5$ plateau is observed in the lowest Landau level, where the Coulomb interaction is more long-ranged as compared to $\hat V_1^{\text{2bdy}}$. Thus apart from the confining potential, insertion of the fluxes and the creation of quasiholes will cost negative amount of energy, i.e. $\epsilon_{N,\alpha}<0$. In the long wavelength limit if we ignore the non-linearity of the confining potential (i.e. $c_2=0$), then we have $\tilde\epsilon_{N,\alpha}<0$ and $\tilde g\left(N,\beta,c_2\right)>\tilde g\left(N,0,0\right)$. A higher density of state leads to an increase of the edge heat capacity, we thus expect $\kappa>\frac{8}{5}\frac{\pi^2k_B^2T}{3h}$. In contrast, stronger confining potential (e.g. sharper edge of the 2DEG in the experiments) generally leads to larger $c_2$ and reduced effective density of states, leading to the suppression of the thermal Hall conductivity. These are some qualitative behaviours we can predict about the experiments based on the simple analysis here, and a more detailed calculations will be presented elsewhere.

\subsubsection{Quasihole tunnelling and shot noise}

In addition to the thermal Hall conductance, we can also explore the topological nature of the quantum Hall fluid at $\nu=2/5$ by looking at charge tunnelling between counter-propagating edges at the quantum point contact (QPC). In the composite fermion picture, quantum fluid consists of two CF levels in analogy to the two LLs of the IQHE, contributing to the two tunnelling channels at the QPC. Like the IQHE, the outer channel has full transmission since they are further apart, and backscattering mainly comes from the inner channel, with the charge carriers each carrying the charge of $q=e/5$. This can be extracted from the relationship of $S=2qI_B$, where $S$ is the spectral function of the shot noise, and $I_B$ is the backscattering current\cite{umansky,umansky1}. 

While the tunnelling of $e/5$ charge carriers have been confirmed in a number of experiments, it has also been discovered at very low temperature, the tunnelling charge is $2e/5$ instead of $e/5$\cite{umansky1}. This rather interesting phenomenon illustrates the richness of edge dynamics at $\nu=2/5$, that cannot be readily explained using the composite fermion theory. Here we show this phenomenon can be naturally explained by looking into the dynamics of the Gaffnian quasiholes, which are non-interacting with $\hat H_{\text{gf}}$, but are no longer the case with realistic interactions.

\begin{figure}
\includegraphics[width=\linewidth]{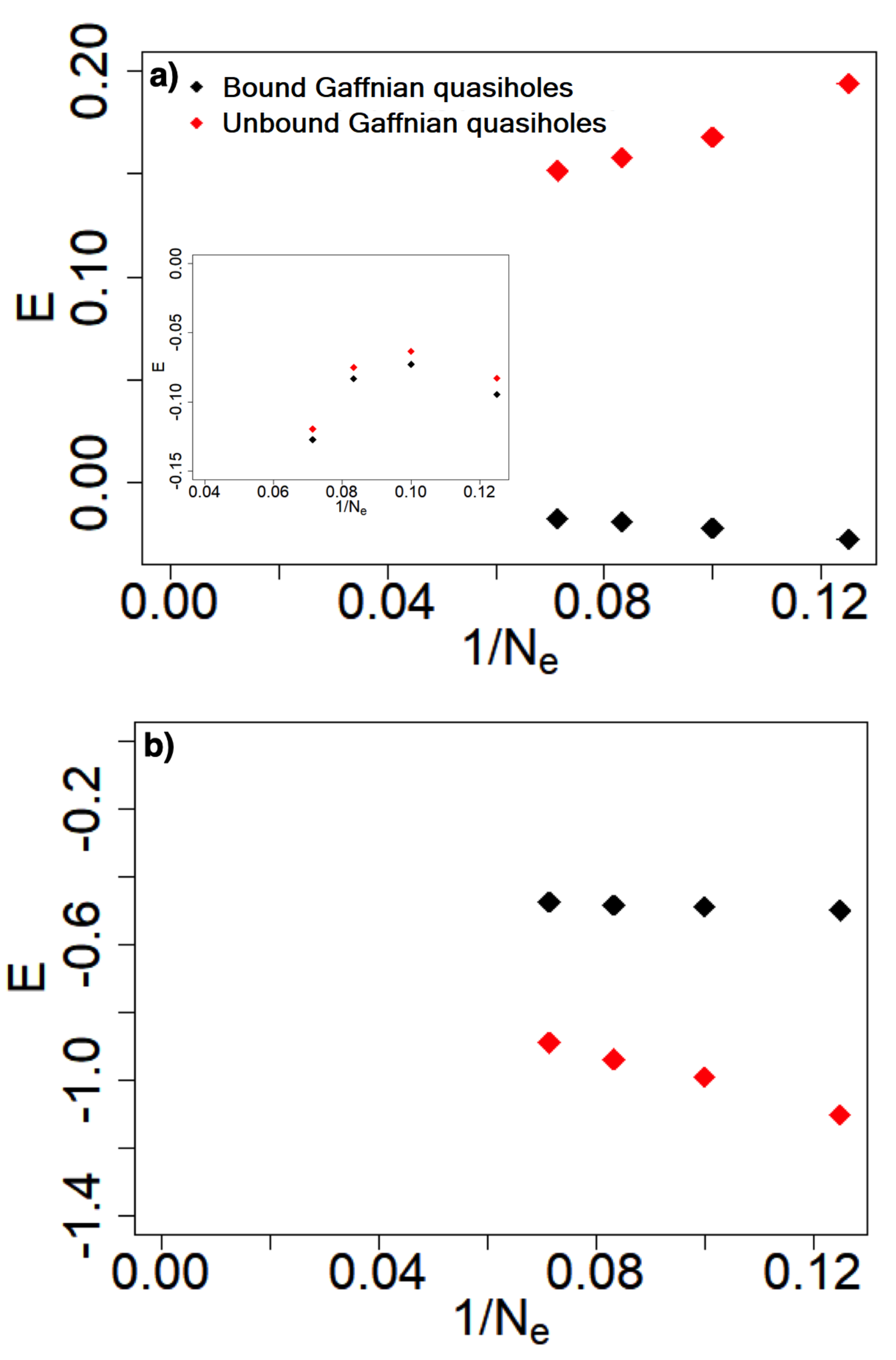}
\caption{The energy cost of creating a pair of bounded (black plot) and unbounded (red plot) Gaffnian quasiholes, after inserting one magnetic flux to the ground state. The x-axis is the inverse of the system size. a). $\hat V_1^{\text{2bdy}}$ interaction; b). $\hat V_3^{\text{2bdy}}$ interaction. The inset shows the same plots with $\hat V_{\text{LLL}}$ Coulomb interaction, with greater finite size effects. For both $\hat V_1^{\text{2bdy}}$ and $\hat V_{\text{LLL}}$ interactions, the unbounded pair of quasiholes has higher variational energy.}
\label{plot3}
\end{figure} 

If we insert one magnetic flux to the Gaffnian ground state and create two Gaffnian quasiholes, the states from the following two root configurations are degenerate with $\hat H_{\text{gf}}$:
\begin{eqnarray}
&&\textsubring{}\textsubring{0}1100011000\cdots 1100011\label{gg1}\\
&&\textsubring{}1010010100\cdots 10100101\textsubring{}\label{gg2}
\end{eqnarray}
where in Eq.(\ref{gg1}) we have two Gaffnian quasiholes forming a bound state with charge $2e/5$, piled at the north pole, in the $L_z=N_e/2$ sector. In contrast, Eq.(\ref{gg2}) is the state with two unbounded Gaffnian quasiholes, one at the north pole and the other at the south pole, in the $L_z=0$ (for $N_e$ even) or $L_z=1$ (for $N_e$ odd) sector. Here we have a Gaffnian quasihole when for five consecutive orbitals, we have one (instead of two) electron, as determined by the admission rule for the Gaffnian ground state. For quasiholes at the north or south pole, the number and the location of the quasiholes can be determined by the position of the inserted flux and the symmetry of the root configuration. We can also apply the admission rule by embedding the root configuration in the ground state fluid (e.g. for Eq.(\ref{gg2}), the quasiholes can be located with the admission rule for the configuration of $\cdots11000110001010010100\cdots101001010001100011\cdots$). With a realistic interaction, we can evaluate their corresponding variational energy to determine if it is more energetically favourable for the two quasiholes to be bounded or unbounded. 

\begin{figure}
\includegraphics[width=\linewidth]{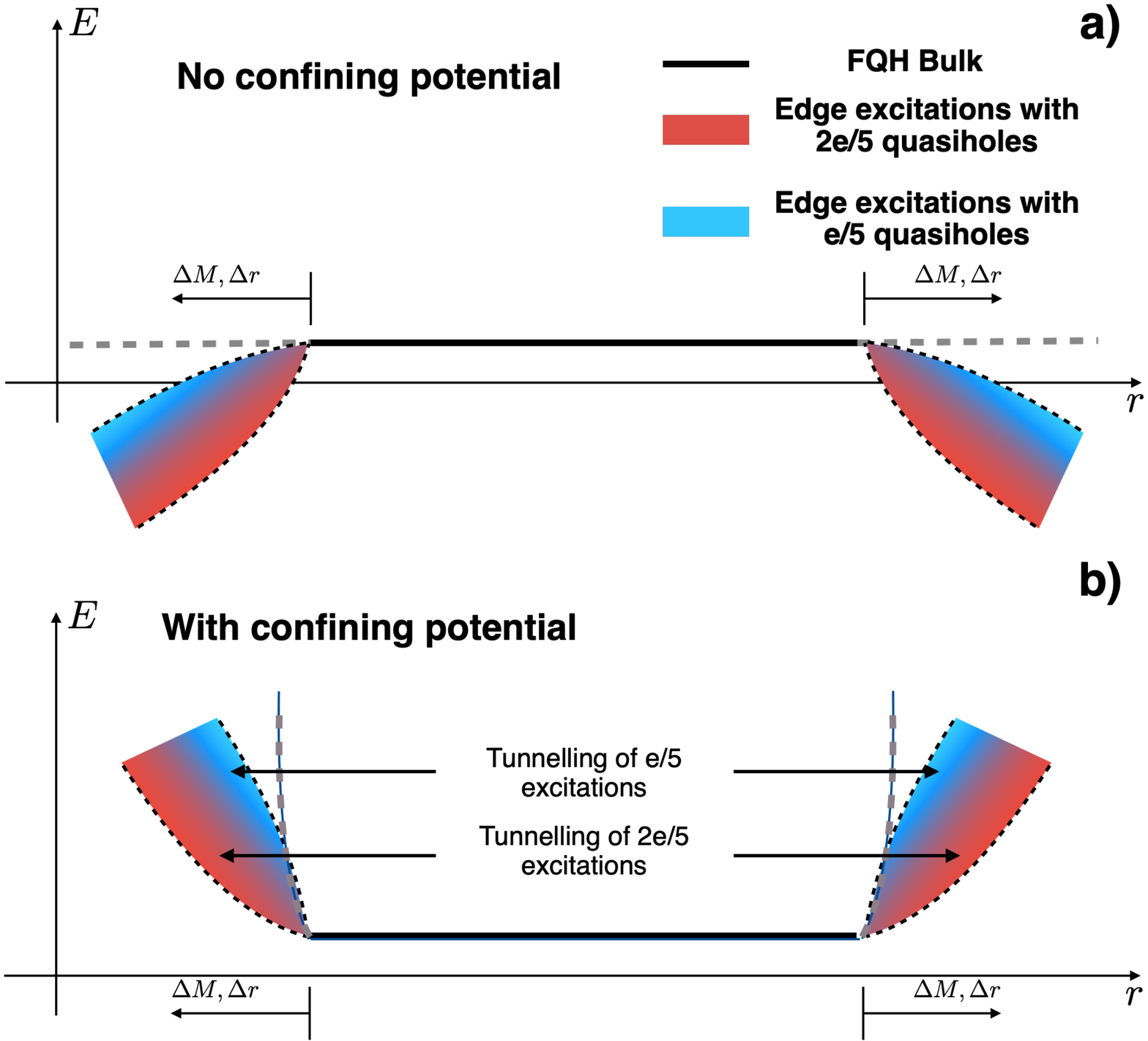}
\caption{Schematic distribution of the $e/5$ and $2e/5$ density at the edge of $\nu=2/5$ FQH fluid, based on the numerical analysis with short range interactions. a). Without confining potential; b). With confining potential.}
\label{fig3}
\end{figure}

From Fig.(\ref{plot3}), we can see from finite size analysis that $\hat V_1^{\text{2bdy}}$ prefers bound quasihole states, while $\hat V_3^{\text{2bdy}}$ prefers unbound quasihole states. Realistic interactions such as LLL Coulomb interaction (i.e. $\hat V_{\text{LLL}}$) is known to be quite close to $\hat V_1^{\text{2bdy}}$, we thus expect it to prefer bound quasihole states as well, as supported by numerical evidence. This also implies the two Gaffnian quasiholes can pull away from each other at finite temperature. Given that the energy difference between the bound and unbound quasihole states seems quite small with $\hat V_{\text{LLL}}$, in realistic samples their separation can be quite large, leading to tunnelling of single-quasihole from one edge to another if the QPC is narrow.

Thus the tunnelling at the QPC can be illustrated in Fig.(\ref{fig3}). At the same temperature, there will always be a higher density of bound quasiholes with charge $2e/5$, as compared to (loosely) unbounded ones carrying the charge of $e/5$. Let the density of the $2e/5$ quasiholes available for tunnelling be $n_{2e/5}$, and the density of the $e/5$ quasiholes available for tunnelling be $n_{e/5}$. We thus have:
\begin{eqnarray}
\frac{n_{e/5}}{n_{2e/5}}\sim e^{-\beta\delta E}
\end{eqnarray}
where $\delta E$ is the characteristic energy difference between bounded and unbounded Gaffnian quasiholes. On the other hand as shown in Fig.(\ref{fig3}), the $e/5$ excitations, being at higher energy, has shorter tunnelling distance. In general the tunnelling amplitude, given by the overlap of the (localised) edge excitations, is suppressed exponentially by the state separation. Thus at low temperature, there are predominantly $2e/5$ excitations at the edge, leading to the shot noise experiment detecting the quantised charge of $2e/5$. As temperature increases, a substantial amount of $e/5$ excitations are present, which dominates the tunnelling process since their tunnelling amplitude is much larger as compared to the $2e/5$ excitations, when the quantum point contact is narrow. Thus there will be a cross-over as temperature increases, and the shot noise measurement will detect mostly quantised charge of $e/5$. This qualitative argument agrees with the experimental observation quite well, and a detailed quantitative analysis with experimental parameters will be performed elsewhere.

\subsection{The Haffnian physics at $\nu=1/3$}

We have already shown that $\hat H_{\text{hf}}$ is gapless in the thermodynamic limit with microscopic arguments that are not readily applicable to the Gaffnian model Hamiltonian. One can then wonder if it is possible to perturb $\hat H_{\text{hf}}$ so that we can have a gapped ground state inheriting some of the topological properties of the Haffnian model state, like what we have shown for the Gaffnian phase. From Fig.(\ref{fig2}), however, we can see that any small perturbation by short range interactions likely leads to level crossing between $\mathcal H^H_{\text{topo}'}$ and $\bar{\mathcal H}^H_{\text{topo}'}$ in the thermodynamic limit. For interactions dominated by $\hat V_1^{\text{2bdy}}$, the ground state and low-lying excitations at $N_o=3N_e-4$ are Laughlin quasielectrons. Increasing $\hat V_3^{\text{2bdy}}$ pushes up the Laughlin quasielectrons, but the low-lying excitations are dominated by $\bar{\mathcal H}_{\text{topo}}^M$, which is contained in $\bar{\mathcal H}^H_{\text{topo}'}$. Thus in simple realistic systems, we do not expect gapped ground state that are topologically equivalent to the Haffnian state, in contrast to the Gaffnian case.

This however does not rule out the possibility of such a gapped phase. If we assume that $\hat H_{\text{hf}}$ gaps out $\bar{\mathcal H}_{\text{topo}}^M$, which is reasonable, we then need a local Hamiltonian to gap out both $\mathcal H_{\text{topo}'}^M$ and $\mathcal H_{\text{topo}'}^G$ for $N_o\le 3N_e-4$, for us to have a gapped phase topologically equivalent to the Haffnian model state, and thus distinct from the Laughlin phase. The topological phase transition from this Haffnian like phase to the Laughlin phase is also accompanied  by the fractionalisation of the Laughlin quasiholes into ``Haffnian" quasiholes\cite{yangbo4} carrying charge of $e/6$. Given that $\mathcal H_{\text{topo}}^L\in\mathcal H_{\text{topo}}^G$, each Laughlin quasihole can be understood as a bound state of two Gaffnian quasiholes. In the Laughlin phase, the bound state is energetically favourable. Pulling the two Gaffnian quasiholes apart costs an energy that is proportional to the distance between them. In the Haffnian-like phase, the unbounded Gaffnian quasiholes are energetically favourable and they emerge as ``Haffnian" quasiholes (note that each Gaffnian quasihole carries the charge of $e/6$ at $\nu=1/3$, and $\mathcal H_{\text{topo}}^H\in\mathcal H_{\text{topo}}^G$). 

If no Haffnian-like gapped phase is possible for any local Hamiltonian, then the only known topological phase at $\nu=1/3$ for a single fermion species is the Laughlin phase. Even in this case, $\mathcal H_{\text{topo}}^H$ can still play a relevant physical role. It is the Hilbert space of the quadrupole excitations of the Laughlin phase, and there are both numerical and experimental evidence that the quadrupole excitations can go soft in the Laughlin phase\cite{yangbo3,yangbo4,xia,fu,maciejko,sondhi}. As long as the charge gap is maintained, the quantum phase still has a robust Hall conductance plateau, though finite temperature transport can be modified by the quadrupole excitations, leading to the so-called nematic FQHE phase. If the quadrupole excitations become gapless, this could imply that the Laughlin state is compressible and degenerate with the Haffnian state, which is a uniform gas of quadrupole excitations. Thus the likely scenario in the experiment is a small energy gap of the quadrupole excitation as compared to the finite temperature. The Laughlin state is still incompressible due to the presence of the charge gap, and that the Haffnian state has an extensive energy gap.

Another possible scenario is for the quadrupole excitation gap to scale as $\sim 1/N_e$, so it becomes gapless in the thermodynamic limit. There is now a finite variational energy gap of the Haffnian model state, which contains $\sim N_e$ number of quadrupole excitations. More interesting there will be a mode with linear dispersion, with the energy scaling linearly with the number of quadrupole excitations. Given that the gaplessness of the quadrupole gap leads to the nematic FQH phase, this linear mode has been described in the effective field theory as the gapless nematic Goldstone mode\cite{yangbo4,maciejko}. One should note that in the thermodynamic limit, a single quadrupole excitation does not have any density modulation, which is unlike the dipole excitations at large momenta. Thus indeed this linear dispersion in the long wavelength limit does not come from the density fluctuation, but from the spatial modulation of the nematic director. It would be interesting to see if this Goldstone mode can be measured experimentally, as it is not clear if it can be realised with realistic interactions.

\section{Summary and discussions}\label{summary}

Using the null space of model Hamiltonians as the preferred degrees of freedom, in this work we argue that the Gaffnian and Haffnian state (as well as their quasihole states) have rich physical properties that can play interesting roles in familiar and exotic gapped FQH systems. The model Hamiltonians of the Gaffnian and the Haffnian state may be gapless in the thermodynamic limit, and we have provided in this work a microscopic argument that the Haffnian model Hamiltonian is indeed gapless against Laughlin quasielectron excitations. On the other hand, the quasihole subspaces are still spanned by well-defined microscopic many-body wavefunctions. In particular, we show at $\nu=2/5$, the Gaffnian quasihole subspace can explain many, if not all, of the topological and non-universal behaviours observed in the experiments with the introduction of the realistic Hamiltonian. This reinforces the previous notion that the Gaffnian formalism and the composite fermion picture are describing the two sides of the same coin\cite{yang1,toke} at $\nu=2/5$. With realistic interaction there is no bulk-edge correspondence. The Gaffnian ground state and the Jain ground state are argued to be topologically equivalent, while the Gaffnian quasihole manifold is split into bands with Coulomb interactions. The dynamics of Gaffnian quasiholes can explain the shot noise and tunnelling experiments, and its prediction of the dependence of the thermal Hall conductance with different tuning parameters can also be checked in experiments.

Both the Gaffnian and Haffnian quasiholes play important roles in the low lying excitations of the Laughlin phase. The quadrupole excitations, which are neutral, are made of Haffnian quasiholes. In contrast, the dipole and quasielectron excitations are made of Gaffnian quasiholes. The energetic competitions between the Haffnian and Gaffnian quasiholes thus give a unifying description of the dynamics of the Laughlin phase at the finite temperature. These include the nematic FQH phase\cite{yangbo3,yangbo4,xia,fu,maciejko,sondhi}, which is a topological phase with non-trivial geometric properties, as well as potential fractionalisation of the Laughlin quasiholes at finite temperature\cite{yangbo4}. We also show that with realistic two-body interactions, there is generally no Haffnian-like ground state similar to the case of the Gaffnian state. This is because the Haffnian quasiholes have very high variational energy as compared to other sub-Hilbert spaces with known realistic interactions. It is still interesting to explore if there exists local Hamiltonians that gaps out all other sub-Hilbert spaces from the Haffnian quasiholes, so as to realise an incompressible Haffnian-like phase with a distinct topological shift as compared to the Laughlin phase at $\nu=1/3$.

It is also important to understand how the results in this work reconciles with the arguments from the effective CFT description, regarding the relevance of the Gaffnian and Haffnian states to gapped FQH phases. We would first like to note that strictly speaking, the results proposed in this work do not contradict the CFT arguments. In those arguments, the gaplessness of the model Hamiltonians for the Gaffnian and Haffnian states requires the fundamental assumption of the conformal invariance of their respective null spaces. Realistic interactions that breaks the conformal symmetry can still retain topological properties of some (may not be all) of those from the model Hamiltonians. Moreover, the CFT arguments do not prevent the Gaffnian and Haffnian quasiholes to be the useful degrees of freedom for low lying excitations of other incompressible FQH phases.

Moreover, we have shown an alternative derivation of the conformal invariance of model Hamiltonian null spaces from the microscopic wavefunctions. In contrast to the effective theories, this derivation shows explicitly how conformal invariance is obeyed by the Hilbert space in the thermodynamic limit, and how the conformal dimension and central charge emerge from the many-body wavefunctions themselves. The delicate structure of the conformal invariance, which we reveal by focusing on the null spaces of three-body pseudopotential interactions, shows that there could be a ``simple" way of manifest conformal symmetry from the microscopic perspective. The negative conformal norm that can be computed microscopically does not lead to unphysical quantum mechanical behaviours of the many-body wavefunctions. It would be interesting to see how different CFT descriptions of the FQH edges are related to each other, which can potentially give us a deeper understanding of how CFT reveals the dynamical properties of both the bulk and edge of the FQH systems.

We end this section with a number of detailed predictions based on the analysis in this work, related to the thermal Hall conductance and the shot noise/quasihole tunnelling experiments. At filling factor $\nu=2/5$ in the LLL, we predict the coefficient of the thermal Hall conductance $\kappa$ to be non-universal and bounded between $\frac{8}{5}$ (with exact Gaffnian quasihole degeneracy) and $2$ (as predicted by the composite fermion theory). With more short ranged interaction (e.g. greater sample thickness or with screening), smaller quasihole creation energy will generally lead to smaller $\kappa$. Non-linear confinement potentials at the edge will also reduce the effective density of state and thus reduce $\kappa$, and we expect that to be the case with a sharper edge. 

The tunnelling experiments at $\nu=2/5$ will involve quasparticles of both $e/5$ and $2e/5$ charge (the latter can be considered as a bound state of two Gaffnian quasiholes). At very low temperature only $2e/5$ quasiholes will be present for the tunnelling. At higher temperature, the edge excitations will consist of a mixture of $e/5$ and $2e/5$ quasiholes, and the former has shorter tunneling distance. Thus for narrow quantum point contact (QPC), at higher temperature the tunnelling can predominantly involve $e/5$ quasiholes. For wider QPC, however, the tunnelling amplitude for $e/5$ quasiholes will be less dominant, and there will be no clean experimental signals of a particular quasihole charge, and we expect some averaged quantities between $e/5$ and $2e/5$ from experiments involving tunnelling between counter-propagating edge currents.

At $\nu=1/3$, the role of Haffnian quasiholes and the experimental raminifcations are mostly predicted in Ref.\cite{yangbo4}. We do not expect a gapped FQH phase with a topological shift $S=-4$ with the LLL or SLL interaction, though this phase is not ruled out in principle. In contrast, there can be a finite temperature phase transition of the quasihole manifold, especially near the nematic FQH phase and at the edge of the Hall plateau (when there is a relatively large quasihole density). This is the remnant of the ``Haffnian phase" with non-Abelian $e/6$ quasiholes that cannot be fully realised due to the gaplessness of the Haffnian model Hamiltonian. Such a phase transition, and the fractionalisation of the $e/3$ Laughlin quasiholes into $e/6$ quasiholes, can in principle be detected by single-electron tunnelling experiments, or the shot noise/inteferometry experiments. Above the crtical temperature for the phase transition, we also expect the thermal Hall conductance coefficient no longer quantised at $\kappa=1$.

\begin{acknowledgments}
I thank Steven Simon for very helpful discussions, and Ajit. C. Balram for the helpful discussions on the Jain states. This work is supported by the Singapore National Research Foundation (NRF) under NRF fellowship award NRF-NRFF12-2020-0005.
\end{acknowledgments}

\end{document}